\documentclass[journal]{IEEEtran}
\usepackage{ulem}
\ifCLASSINFOpdf
\else
  \usepackage[dvips]{graphicx}
\fi
\usepackage[cmex10]{amsmath}

\usepackage{stfloats}

\usepackage{bm}
\begin{document}
%
\title{Multi-hop Analog Network Coding: An Amplify-and-Forward Approach}
%
%
%

\author{Binyue~Liu,
        and~Ning~Cai,~\IEEEmembership{Senior~Member,~IEEE}
\thanks{This work was supported by by grants from the National
Natural Science Foundation of China (60832001). The material in this
paper was presented in part at IEEE International Symposium on Information Theory, St. Petersburg, Russia, Aug. 2011,
the 7th Asia Europe Workshop on Concepts in Information theory, Boppard, Germany, Jul. 2011, and
2012 Information Theory and Applications Workshop, San Diego, Feb. 2012.}
\thanks{The authors are with the Key Lab. of ISN, Xidian University, Xi'an, China e-mail: \{liuby, caining\}@mail.xidian.edu.cn.}
}
\maketitle

\begin{abstract}
In this paper, we study the performance of an amplify-and-forward (AF) based \emph{analog network coding} (ANC) relay scheme in a multi-hop wireless network under individual power constraints.
In the first part, a unicast scenario is considered. The problem of finding the maximum achievable rate is formulated as an optimization problem.
Rather than solving this non-concave maximization problem, we derive upper and lower bounds for the optimal rate.
A cut-set like upper bound is obtained in a closed form for a layered relay network.
A pseudo-optimal AF scheme is developed for a two-hop parallel network, which is different from the conventional scheme with all amplification gains chosen as the maximum possible values.
The conditions under which either the novel scheme or the conventional one achieves a rate within half a bit of the upper bound are found.
Then we provide an AF-based multi-hop ANC scheme with the two schemes for a layered relay network.
It is demonstrated that the lower bound of the optimal rate can asymptotically achieve the upper bound when the network is in the generalized high-SNR regime.
In the second part, the optimal rate region for a two-hop multiple access channel (MAC) via AF relays is investigated.
In a similar manner, we first derive an outer bound for it and then focus on designing low complexity AF-based ANC schemes for different scenarios.
Several examples are given and the numerical results indicate that the achievable rate region of the ANC schemes can perform close to the outer bound.
\end{abstract}

\begin{IEEEkeywords}
Amplify-and-forward (AF), analog network coding (ANC), wireless networks, multiple access.
\end{IEEEkeywords}

%
\IEEEpeerreviewmaketitle

\section{Introduction}
%
%
%
%
\IEEEPARstart{S}{ince} the introduction of the amplify-and-forward (AF) relay scheme, it has been studied in the context of cooperative
 communication [1]-[4]. It is an interesting technique from the practical standpoint because the complexity and cost of relaying,
always an issue in designing cooperative networks, is minimal for AF relay networks. As the simplest coding scheme, AF is also used to
estimate the network capacity of relay networks. Obviously, the achievable rate of AF scheme can be viewed as a lower bound to the
network capacity. In addition to its simplicity, AF is known to be the optimal relay strategy in many interesting cases [5]-[7].
Gastpar and Vetterli [7] have shown that for a two-hop network with AF relays, the cut-set bound of the network capacity can be
achieved in the limit of a large number of relays.

Network coding [9] is a novel and promising design paradigm for wired communication networks. As opposite to the conventional routing operation, network coding allows the intermediate relays processing the received packets to reduce the amount of transmissions and thus improves the total throughput of the network. Li \emph{et al.} [8] have shown that linear network coding can achieve the multicast capacity [9] in a noiseless network. This result indicates that to send out a linear combination of the incoming packets at each node is sufficient to obtain the optimal capacity performance. Linear network coding has also been studied in [10] from an algebraic perspective. Each destination node effectively obtains source information multiplied by a transfer matrix consisting of global encoding kernels on the incoming edges of it, and can recover the original data provided that the matrix is invertible.

Applying the principle of network coding to wireless communication networks has recently received tremendous attention from
the research community. AF relay scheme allows one to exploit the broadcast nature of the wireless medium and to introduce the
concept of network coding into physical layer. Katti \emph{et al.} [11] have studied an AF-based analog network coding (ANC) scheme. As opposite to the traditional approach of wireless communications, analog network coding fully use the interference rather than avoiding it. This technique significantly improves the network throughput in many scenarios. Since then, many works have been focused on the design of ANC relay schemes both for one-way and two-way relay channels [12]-[17]. In [15] Mari\'c \emph{et al.} have studied a multi-hop ANC scheme for a layered relay network under the individual power constraint. With each relay node amplifying the received signal to the maximum possible value, the achievable rate is shown to approach the network capacity in the high-SNR regime. Later such result is extended to the generalized high-SNR regime as defined in [16]. The results in [16] have shown the effectiveness of ANC in such generalized scenario. Recently, Agnihotri \emph{et al.} [17] propose an iterative algorithm to obtain the optimal ANC scheme for a layered relay network with "Equal Channel Gains between Adjacent Layers (ECGAL)" property in general SNR regime.

To employ ANC in a wireless network with AF relays, we are especially interested in deriving the optimal achievable rate. There have been many results for this problem [18]-[20]. For a two-hop parallel relay network, Mari\'c and Yates [18] have found the optimal AF relay scheme in closed-form along with the maximum achievable rate under the sum power constraint. A more general result is obtained by Gomadam and Jafar [19]. A case when all the relay nodes introduce the correlated Gaussian noises is considered. They have found the optimal AF relay scheme for this scenario and investigated the influence of the correlation between noises on the end-to-end performance. However, the optimal AF scheme for a general network where both the topology and the operation regime of it have no limitation, is still unknown, even in the case of a two-hop parallel relay network under the individual power constraints. Agnihotri \emph{et al.} [20] have provided a framework to compute the maximum achievable rate with AF schemes for a class of general wireless relay networks, which casts the problem as an optimization problem. The similar idea also appears in an independent work [16]. Unfortunately, the optimization problem in general case is hard to be solved.

To employ AF-based ANC scheme in a wireless network, there are still many important and interesting problems remaining to be solved. We list some of them as follows.
\begin{itemize}
\item Problem 1: The cut-set bound is usually used to justify the performance of different relay schemes. However, it is not a tight upper bound for the AF achievable rate in most cases because the relay nodes are only allowed to do linear operations. To derive a tight upper bound for the achievable rate of a network with AF relays is an issue of great importance since the optimal rate is always hard to be obtained.
\item Problem 2: Is it always optimal for relay nodes to amplify the received signals to the maximum possible values under individual power constraint? Intuitively, for different scenarios there may exist different AF schemes that have a better end-to-end throughput performance than any others. Then how to characterize them?
\item Problem 3: A multi-user scenario is also an interesting topic [21]. For a two-hop MAC with AF relays, how many AF schemes shall we use to obtain the entire achievable rate region? Obviously, it is infeasible from the practical standpoint for the relay nodes to store all the AF schemes. Hence, it is worth investigating the tradeoff between the performance and the complexity.
\end{itemize}

The main results of this paper give partial answers to the above questions and are summarized as follows:
\begin{itemize}
\item In this paper, we pursue two related objectives. The first one is to investigate the optimal AF-based ANC scheme for a multi-hop unicast relay network. Assuming each relay node has a transmitting power constraint, we derive an upper bound of the optimal achievable ANC rate in a closed form for a layered relay network. The idea behind our method is similar to the technique used to derive the cut-set bound. Thus we call it a cut-set like upper bound. A novel AF scheme for a two-hop parallel relay network is proposed, which is different from the conventional scheme with all relay nodes amplifying the received signal to the maximum possible value. We determine the different conditions under which either the novel scheme or the conventional one has a better performance to approach within $1/2$ bit to the upper bound and thus the optimal rate. Based on such observations, a mixed AF-based multi-hop ANC scheme is proposed. When the network is in the generalized high-SNR regime [16], we demonstrate that the lower bound of the achievable rate can asymptotically achieve the upper bound. Furthermore, we point out that the result obtained in [15] can be viewed as a special case of the result obtained in our paper.
\item The second objective of this work is to extend the ANC scheme to the multiuser case. S. A. Jafar \emph{et al.} [21] have shown the optimal rate region of the two-hop parallel AF network under a sum power constraint. We find that to assume the relay nodes under individual power constraints makes the problem more complicated. It is observed that for a specific AF scheme, the network reduces to a conventional MAC. Each Gaussian capacity in the rate set of the MAC is equivalent to the capacity of a two-hop parallel unicast network. With the similar idea, we first derive two outer bounds for the optimal rate region. The structures of them are fully characterized. Then a dynamic ANC scheme is proposed. According to different network settings, the number of the AF schemes stored in the relay nodes varies and with finitely many schemes a time-sharing inner bound is obtained. The tradeoff between the storage space and the performance is decided for the practical purpose. Finally, we illustrate three ANC schemes. The numerical result shows that the inner bound of each ANC scheme is close to the respective outer bound in different scenarios.
\end{itemize}

\emph{Notation}: Scalars are denoted by lower-case letters, e.g., $x$,
and bold-face lower-case letters are used for vectors, e.g., $\bf{x}$,
and bold-face upper-case letters for matrices, e.g., $\bf{X}$. In addition,
$trace(\bf{X})$, $|\bf{X}|$, ${\bf{X}}^{T}$ and ${\bf{X}}^{-1}$ denote the trace, determinant, transpose and inverse matrix of $\bf{X}$, respectively, and
$diag(x_1, \cdots ,x_n)$ denotes a block-diagonal square matrix
with $x_1, \cdots ,x_n$ as the diagonal elements. ${\bf{X}}(i, j)$ denotes the $(i, j)$-th element of $\bf{X}$, and $rank({\bf{X}})$ denotes the rank of ${\bf{X}}$. $||{\bf{x}}||$ denotes the Euclidean norm of a vector ${\bf{x}}$. $E\left[  \cdot  \right]$ is the expectation operation. $log(\cdot)$ denotes the
logarithm in the base 2 and $ln(\cdot)$ denotes the natural logarithm. We use $conv \left( \cdot \right)$ and $col \left( \cdot \right)$ to denote the convex hull and the closure of a set.

\section{Network Model}
In the first part of this paper, we analyze a multi-hop Gaussian relay network with a single source-destination pair $\left( {S,D} \right)$, which is represented by a directed graph ${\cal{G}}=({\cal{V}},{\cal{E)}}$ depicted in Fig. 1. Assume each non-source node $k$ introduces a Gaussian noise, all the relay nodes work in a full-duplex mode, and there is no circle path in the network. At instant $n$, the channel output at node $k$ is a linear combination of all noisy signals transmitted from its upstream nodes and the Gaussian noise introduced by itself and can be expressed as
\begin{equation}
y_k\left[ n \right]  = \sum\limits_{j \in {\cal{V}}\left( k \right)} {h_{j,k} x_j \left[ n \right] + z_k\left[ n \right] },
\end{equation}
where $h_{j,k}$ denotes the channel gain from node $j$ to node $k$, ${\cal{V}}(k)$ represents neighboring nodes of node $k$ which have one-hop path to node $k$, $x_j\left[ n \right]$ is the transmitted signal at node $j$, and  $z_k\left[ n \right]$ is a sequence of independently and identically distributed (\emph{i.i.d.}) Gaussian noises with zero mean and variance $1$. All the channel gains are supposed to be fixed positive real-valued constants known through the network for the scope of the present paper.

\begin{figure}
  \centering
  \includegraphics[width=2.4in]{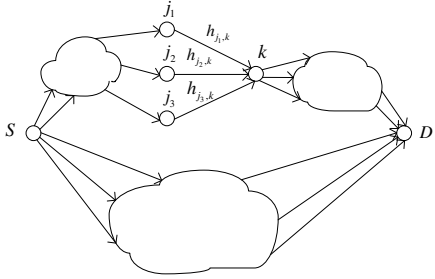}
  \caption{Wireless relay network with single source-destination pair}
\end{figure}

A natural assumption is that there exists a transmitting power constraint $P_k^{Up}$ at each node $k$ such that
\begin{equation}
E\left[ {x_k^2 } \right] \le P_k^{Up}.
\end{equation}
Each network node performs analog network coding via amplify-and-forward relay scheme. Assume the relay nodes operate instantaneously as in [12], that is, the relay nodes amplify and forward their input signals without delay.
\begin{equation}
x_k \left[ n \right] = \beta _k y_k\left[ n \right],
\end{equation}
where the amplification gain $\beta _k$ is chosen such that the power constraint (2) is satisfied. Actually the possible system instability resulting from this assumption is avoided by a "buffering and subtracting" technique as observed in [20]. So, with this assumption, the time index $n$ then can be omitted in the sequel for the sake of brevity. Through a relaying path from node $j$ to node $k$, the signal $x_j$ is multiplied by the amplification gains and the channel gains along this directed path. Since the multi-hop ANC scheme takes more of the network coding approach, the corresponding coding coefficients are defined as follows.
\newtheorem{definition}{Definition}
\begin{definition}[Local Encoding Coefficient]
Let $e(i,j)$ and $e(j,k)$ be the input and output channels of node $j$. We call
\begin{equation}
\alpha _{e(j,k) }  = \beta _j h_{e(j,k) }
\end{equation}
the local encoding coefficient of pair $\left( {e(i,j) , e(j,k) } \right), i \in {\cal{V}}(j)$, where, in particular, $
\beta _S  = 1$.
\end{definition}

\newtheorem{definition2}{Definition}
\begin{definition}[Global Encoding Coefficient]
 We call
\begin{equation}
f_{j,k}  = \sum\limits_{\left\{ {\cal{E}}(j,k) \right\}} {\prod\limits_{e(m_1,m_2) \in {\cal{E}}(j,k)} {\alpha_{e(m_1,m_2) } } }
\end{equation}
the global encoding coefficient from node $j$ to node $k$, where ${\cal{E}}(j,k)$ represents the set of channels appearing in a relay path between the two nodes.
\end{definition}

The global encoding coefficient defined here can be interpreted as the equivalent path gain between node $j$ and $k$.
The relationship between the local and global coefficients is given as follows.
\begin{equation}
f_{S,k}  = \sum\limits_{j \in {\cal{V}}\left( k \right)} \alpha _{e(j,k) } f_{S,j}
\end{equation}

\section{Upper Bound to ANC Rate}
\subsection{Sufficient Condition of Power Constraint}
We first derive the expression of the achievable rate for a unicast network with a specific ANC scheme. Then the problem of finding the maximum achievable rate is formulated as an optimization problem.  In this subsection, we focus on capturing the feasible domain of each amplification gain according to the power constraint at each relay node.

According to the assumption that each non-source node introduces a Gaussian noise, each of them can be viewed as a source with the noise as the input. Therefore, by the linear operation of AF relay scheme, the original network can be considered as a linear combination of several unicast subnetworks and the source of each subnetwork is either the source node $S$ or the non-source node. Therefore, with the global encoding coefficient (5), the received signal at the destination is expressed as
\begin{align}
y_D & = f_{{\rm{S}},D} x_{\rm{S}}  + \sum\limits_{\left\{ {j } \right\}} {f_{j ,D} z_{j } }  + z_D  \nonumber \\
  &= x_{S,eq}  + z_{eq},
\end{align}
where $x_{S,eq}  = f_{{\rm{S}},D} x_{\rm{S}}$ represents the equivalent signal transmitted from the source node, and $z_{eq}  = \sum\limits_{\left\{ {j } \right\}} {f_{j ,D} z_{j } }  + z_D$ denotes the equivalent noise received at the destination node where the summation is over all relay nodes $\{j\}$. By the assumption that all the Gaussian noises introduced at the non-source nodes are independent, the sum of them is also drawn according to Gaussian distribution. Hence, from the RHS of the second equality above, the relay network employing the ANC relaying scheme can be considered as a point-to-point Gaussian channel. As the well-known result, the source node adopts the Gaussian codebook. The random variable (r.v.) $x_s$ used to generate the codewords is drawn according to $\mathcal{N}\left[ {0,P_s } \right]$ , where $\mathcal{N}\left[ {0,\sigma ^2 } \right]$ denotes normal distribution with
zero mean and variance $\sigma^2$. A sequence of codes containing $\left\lceil {2^{nR} } \right\rceil$ codewords of length $n$ is proposed and it is shown that the error probability goes to zero as $n \to \infty$.

From (4), (5) and (7), both the equivalent signal and noise are related to the amplification gains chosen in the ANC scheme. For the equivalent Gaussian channel, we focus on the choice of the amplification gains to maximize the SNR value at the destination node, and thus the transmission rate $R$. Certainly, this problem can be formulated as a standard optimization problem with the object function being the SNR received at the destination node which is shown as follows.
\begin{equation}
SNR\left( {\left\{ {\beta _k , k \in {\cal{V}} \verb|\| (S,D) } \right\}} \right) = \frac{{f_{S,D}^2 P_S }}{{\sum\limits_{{j \in {\cal{V}} \backslash (S,D)}} {f_{j ,D}^2 }  + 1}},
\end{equation}
which is subject to the power constraints at the relay nodes. Hence, to solve this optimization problem, it is necessary to convert the power constrains into the constraints with respect to the amplification gains first. However, for a multi-hop relay network, the selection of the amplification gain at one node also relies on the selections at its upstream nodes. Therefore, we find it hard to give an equivalent independent constraint for each amplification gain. To simplify the problem, we first provide such an independent constraint. Then we show it suffice for the power constraint. Fortunately, it will be shown that under such constraint, a good multi-hop ANC scheme can be proposed in a special SNR regime defined in the sequel. Several parameters are defined first as observed in [15].
\begin{definition}
When each node j transmits with $P_j^{Up}$ given in (2), the power received at node k is upper bounded by
\begin{equation}
P_{R,k}  = \left( {\sum\limits_{j \in {\cal{V}}\left( k \right)} {h_{j,k} \sqrt {P_j^{Up} } } } \right)^2,
\end{equation}
and the reciprocal of $P_{R,k}$ is represented by
\begin{equation}
\delta _k  = \frac{1}{{P_{R,k} }}.
\end{equation}
\end{definition}

Note that if the network is noiseless, then the received signal power at node $k$ approaches (9) when all the neighboring nodes of it transmit the received signals at the upper bounds of the power constraints (2). Then we extend the sufficient condition of the power constraint in [15] to a general network model described in section II by the following lemma.
\newtheorem{lemma}{Lemma}
\begin{lemma}
In a multi-hop Gaussian relay network, it is sufficient for each relay node $k$ to choose the amplification gain
\begin{equation}
\beta _k^2  \le \frac{{P_k^{Up} }}{{\left( {1 + \delta _k } \right)P_{R,k} }} \buildrel \Delta \over = \left(\beta^{Up}_k\right)^2
\end{equation}
such that the power constraint (2) is satisfied.
\end{lemma}
\begin{IEEEproof}
The proof is given in the appendix A.
\end{IEEEproof}

Note that by Definition 3, the value $P_{R,k}$ and its reciprocal $\delta _k$ are both constants determined by the network settings. Therefore, the constraints for the amplification gains proposed in the above lemma are independent of each other.
\subsection{Upper Bound to ANC Rate}
\begin{figure}
  \centering
  \includegraphics[width=2.4in]{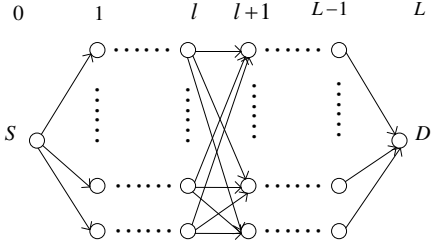}
  \caption{Layered relay network with single source-destination pair}
\end{figure}

In this subsection, we first show a layered relay network depicted in Fig. 2. The source node is assumed to be at layer $0$ and the destination node at layer $L$. The number of relay nodes at layer $l$ is denoted by $n_l$. The transmitted signal vector at layer $l$ is denoted by ${\bf{x}}_l$ and the received signal vector at layer $l+1$ can then be obtained as
\begin{equation}
{\bf{y}}_{l + 1}  = {\bf{H}}_l {\bf{x}}_l +{\bf{  z}}_{l + 1},
\end{equation}
where ${\bf{y}}_{l + 1}$ and ${\bf{  z}}_{l + 1}$ are the received signal vector and the noise vector at layer $l+1$ respectively, and ${\bf{H}}_l$ denotes the channel matrix between these two layers with the element ${\bf{H}}_l \left( {j,k} \right)$ representing the channel gain from node $j$ at layer $l$ to node $k$ at layer $l+1$. Especially, we use ${\bf{h}}_0$ to denote the broadcast channel between the source node and the nodes in the first layer and ${\bf{h}}_{L-1}$ to denote the multiple access channel between the nodes in the $L-1$th layer to the destination node.

Specifically, the global encoding coefficients in such network are given as follows,
\begin{equation}
f_{S,D}  = {\bf{h}}_{L - 1}^T {\bf{B}}_{L - 1}  \cdots {\bf{H}}_1 {\bf{B}}_1 {\bf{h}}_0,
\end{equation}
\begin{equation}
f_{j ,D}  = {\bf{h}}_{L - 1}^T {\bf{B}}_{L - 1}  \cdots {\bf{h}}_{l,j } \beta _{j } ,j  \in {\cal{L}}_l ,l = 1,2 \cdots L - 1,
\end{equation}
where ${\bf{h}}_{l,j }$ represents the $j$th column-vector in ${\bf{H}}_l$, and ${\bf{B}}_l  = {\rm{diag}}\left\{ {\beta {}_1 \cdots \beta {}_j \cdots \beta {}_{n_l }} \right\},j \in {\cal{L}}_l$ where ${\cal{L}}_l$ denotes the node set of layer $l$. From (7), (13) and (14), the signal received at the destination node is given by
\begin{align}
y_D  = {\bf{h}}_{L - 1}^T & {\bf{B}}_{L - 1} \cdots {\bf{H}}_1 {\bf{B}}_1 {\bf{h}}_0 x_s \nonumber\\
 &+ \sum\limits_{l = 1}^{L - 1} {{\bf{h}}_{L - 1}^T {\bf{B}}_{L - 1}  \cdots {\bf{H}}_l {\bf{B}}_l {\bf{z}}_l }  + z_D,
\end{align}
and the SNR function at the destination node (8) results to be
\begin{align}
S&NR\left( {\left\{ {\beta _k ,k \in {\cal{L}}_1,{\cal{L}}_2 \cdots {\cal{L}}_{L-1}} \right\}} \right) \nonumber\\
&= \frac{{E\left[ {\left( {{\bf{h}}_{L - 1}^T {\bf{B}}_{L - 1}  \cdots {\bf{H}}_1 {\bf{B}}_1 {\bf{h}}_0 x_s } \right)^2 } \right]}}{{E\left[ {\left( {\sum\limits_{l = 1}^{L - 1} {{\bf{h}}_{L - 1}^T {\bf{B}}_{L - 1}  \cdots {\bf{H}}_l {\bf{B}}_l {\bf{z}}_l } } \right)^2 } \right] + E\left[ {z_D^2 } \right]}}.
\end{align}

To have an upper bound to the ANC rate, we recall the technique of obtaining the cut-set bound of capacity of a noisy network. One may first assume all channels in the network are noiseless except at one cut and then have an upper bound. By taking the minimum upper bounds over all cuts, the cut-set bound is derived. By applying the same idea to ANC in layered networks, an analysis of an ideal layered relay network is made first. We first fix a layer $l_0$ and assume the Gaussian noises are only introduced at nodes on this layer and other parts in the network are noiseless. By optimizing the SNR function received at the destination node of such network which is formulated by
\begin{align}
SNR_{l_0}& \left( {\left\{ {\beta _k ,k \in {\cal{L}}_1,{\cal{L}}_2 \cdots {\cal{L}}_{L-1}} \right\}} \right)\nonumber\\
&= \frac{{E\left[ {\left( {{\bf{h}}_{L - 1}^T {\bf{B}}_{L - 1}  \cdots {\bf{B}}_1 {\bf{h}}_0 x_s } \right)^2 } \right]}}{{E\left[ {\left( {{\bf{h}}_{L - 1}^T {\bf{B}}_{L - 1}  \cdots {\bf{H}}_{l_0 } {\bf{B}}_{l_0 } {\bf{z}}_{l_0 } } \right)^2 } \right]}},
\end{align}
an upper bound of the optimal ANC rate $R_{opt}$ is derived. Then by traversing $l_0=1,2 \cdots L$, a collection of upper bounds is obtained. To choose the minimum of them, a better upper bound is derived. We draw the conclusion in the following theorem.
\newtheorem{thm}{Theorem}
\begin{thm}[Upper Bound to ANC Rate]
An upper bound of the ANC rate of a Gaussian layered relay network with $L$ layers under the
individual transmitting power constraints is given as follows.
\[
R^{Up}  = \mathop {\min }\limits_{l_0  = 1,2 \cdots L} \frac{1}{2}\log \left( {1 + {\bf{P}}_{R,l_0 }^T {\bf{P}}_{R,l_0 } } \right),
\]
where ${\bf{P}}_{R,l_0}  = \left[ {\sqrt {P_{R,1} }  \cdots \sqrt {P_{R,j} }  \cdots \sqrt {P_{R,n_{l_0} } } } \right]^T ,j
\in {\cal{L}}_{l_0}.$
\end{thm}
\begin{IEEEproof}
The proof is given in the appendix B.
\end{IEEEproof}

We refer to the optimal AF scheme at layer $l_0$ for the ideal network with only the nodes in layer $l_0$ introducing the noises as a "pseudo-optimal amplification scheme".
Note that the idea behind our upper bound is similar to the idea of cut-set bound, we call this upper bound a cut-set like upper bound. As we know, the cut-set bound is an upper bound to the network capacity, where the relay nodes are allowed to do any operations,  including decoding. So we intuitively feel that the cut-set bound may not be tight for the optimal achievable rate of a network with AF relays in general case. To generalize the result of Theorem 1, we hope to find out a more general upper bound result for a non-layered relay network with AF relays. By taking a cut set of the network, we use the notation $\cal{J}$ to denote the nodes on the output sides of the channels in such cut set. If assume only the nodes $j$'s in $\cal{J}$ introduce the Gaussian noises, the corresponding SNR function can be expressed as,
\begin{equation}
SNR_{\cal{J}} \left( {\left\{ {\beta _k, k \in {\cal{V}} \verb|\| (S,D) } \right\}} \right) = \frac{{f_{S,D}^2 P_S }}{{\sum\limits_{j \in {\cal{J}}} {f_{j ,D}^2 } }}.
\end{equation}
Using the same argument as we derive Theorem 1, by optimizing $SNR_{\cal{J}}$, an upper bound of $R_{opt}$ is derived. Then the upper bound of ANC rate in general network is given as
\begin{equation}
R^{Up}= \mathop {\min}\limits_{\cal{J}} \mathop {\max}\limits_{{\beta _k, k \in {\cal{V}} \backslash (S,D) } } \frac{1}{2} \log \left( 1 + SNR_{\cal{J}} \right).
\end{equation}
However, we are unable to give a closed-form of the optimal solution for $SNR_{\cal{J}}$ in general case.

\newtheorem{remark}{Remark}
\begin{remark}
The upper bound of the ANC rate in [15] can be regarded as a corollary of Theorem $1$. Let us revisit the result obtained in [15] in an alternative approach proposed in our paper. Since in the scenario considered in paper [15], all the relay nodes have sufficiently large transmitting powers except the ones at layer $L-1$, the upper bound in Theorem $1$ shows that the minimum is taken when $l_0=L$. The SNR function of an ideal network with $l_0=L$ , i.e., only the destination node introduces the Gaussian noise turns to be
\begin{align}
SNR_L&\left\{ {\beta _k ,k \in {\cal{L}}_1,{\cal{L}}_2 \cdots {\cal{L}}_{L - 1}} \right\} \nonumber \\
&= \frac{{E\left[ {\left( {{\bf{h}}_{L - 1}^T {\bf{B}}_{L - 1}  \cdots {\bf{H}}_1 {\bf{B}}_1 {\bf{h}}_0 x_s } \right)^2 } \right]}}{{E\left[ {z_D^2 } \right]}},
\end{align}
then the corresponding upper bound of the achievable ANC rate is
\begin{equation}
R^{Up} = \frac{1}{2}\log \left( {1 + P_{R,D} } \right).
\end{equation}
Furthermore, from (20) we can see that the noise power received at the destination is independent of amplification gains in this case. Hence, the larger the transmitting powers are, the better the performance of the ANC scheme. With the power constraint, a practical ANC scheme is proposed as observed in [15].
\begin{equation}
\beta _k  = \beta^{Up}_k , k \in {\cal{L}}_l, l = 1,2, \cdots ,L - 1.
\end{equation}
\end{remark}
\newtheorem{remark2}{Remark}
\begin{remark}
In [7], an upper bound to capacity of the relay network is computed using a weaker corollary of cut-set theorem [25, Theorem 14.10.1] and the capacity formula for Gaussian vector channels with fixed transfer function [24]. Applying the similar idea used to verify the corollary, we can also compute the upper bound to capacity of the layered relay network, which can be formulated as an optimization problem as follows.
\begin{equation}
\mathop {\min }\limits_{l_0=1, \cdots , L-1} \mathop {\max }\limits_{p\left(x_j, j \in {{\cal{L}}_{_{l_0  - 1} } }\right) } {\rm{   }}I\left( {{\bf{x}}_{l_0  - 1} ;{\bf{y}}_{l_0 } } \right)
\end{equation}
subject to
\begin{equation}
E\left[ {x_j^2 } \right] \le P_j^{Up} ,j \in {\cal{L}}_{l_0  - 1}
\end{equation}
Let us relax this power constraint to
\begin{equation}
\sum\limits_{j \in {\cal{L}}_{l_0  - 1} } {E\left[ {x_j^2 } \right]}  \le \sum\limits_{j \in {\cal{L}}_{l_0  - 1} } {P^{Up}_j }
\end{equation}
Note that for given $l_0$, (23) is formulated due to the assumption that the nodes in layer $l_0-1$ are the multiple transmitting antennas of the source node and the nodes in layer $l_0$ are the multiple receiving antennas of the destination node. Hence, the transmitting antennas may have any cooperation to encode the message and the multiple receiving antennas may have any cooperation to decode it. Consequently, the corresponding capacity is an upper bound to the network capacity. The solution of the problem defined by (23) and (24) cannot be larger than the solution of the problem defined by (23) and (25). For a specific $l_0$, the latter can be evaluated by using the result in [24].
\begin{equation}
C_{l_0 }= \mathop {\max }\limits_{{{\bf{K}}_{{\bf{x}}_{l_0  - 1} } }} \frac{1}{2}\log \left( {\det \left( {I_{n_{l_0 } }  + {\bf{H}}_{l_0  - 1} {\bf{K}}_{{\bf{x}}_{l_0  - 1} } {\bf{H}}_{l_0  - 1}^T } \right)} \right),
\end{equation}
where ${{\bf{K}}_{{\bf{x}}_{l_0  - 1} } } = E \left[ {\bf{x}}_{l_0-1} {\bf{x}}_{l_0-1}^T \right] $ and the maximum is subject to ${\rm{trace}} \left( {{\bf{K}}_{{\bf{x}}_{l_0  - 1} } } \right) \le \sum\limits_{j \in {\cal{L}}_{l_0  - 1} } {P_j^{Up} }$. Note that the optimal solution is obtained by water-filling. To investigate how far between the cut-set bound and the upper bound derived in Theorem 1, several special cases are taken into account.

When the minimum in Theorem $1$ is obtained by taking $l_0=1$, i.e., the first hop is on the bottleneck in the data transfer, we have the broadcast cut-set bound as the upper bound for capacity.
\begin{equation}
C_{BC}  = \frac{1}{2}\log \left( {1 + {\bf{h}}_0^T {\bf{h}}_0 P_S } \right).
\end{equation}
At this time, the upper bound obtained in Theorem 1 is
\begin{equation}
R^{Up}  = \frac{1}{2}\log \left( {1 + {\bf{h}}_0^T {\bf{h}}_0 P_S } \right),
\end{equation}
which is equal to the former. Hence, in case when $R^{Up}$ is achieved, then the network capacity can be achieved.

When the minimum in Theorem $1$ is obtained by taking $l_0=L$, i.e., the last hop is on the bottleneck of the data transfer, we have the multiple access cut-set bound as the upper bound for capacity.
\begin{align}
C_{MAC}  &= \frac{1}{2}\log \left( {1 + {\bf{h}}_{L - 1}^T {\bf{h}}_{L - 1} \left( {\sum\limits_{j \in {\cal{L}}_{L - 1} } {P_j^{Up} } } \right)} \right) \nonumber \\
&= \frac{1}{2}\log \left( {1 + {\bf{h}}_{L - 1}^T {\bf{h}}_{L - 1} \left( {{\bf{P}}^{Up} } \right)^T {\bf{P}}^{Up} } \right),
\end{align}
where ${\bf{P}}^{Up}= \left[ \sqrt{P^{Up}_1} \cdots \sqrt{P^{Up}_j }\cdots \sqrt{P^{Up}_{n_{L-1}}} \right]$.
At this time, the upper bound obtained in Theorem 1 is
\begin{align}
R^{Up} & =  \frac{1}{2}\log \left( 1 + P_{R,D } \right) \nonumber \\
&= \frac{1}{2}\log \left( {1 + ({\bf{h}}_{L - 1}^T {\bf{P}}^{Up})^2 } \right).
\end{align}
By Schwarz inequality, $R^{Up} \le C_{MAC}$. The equality holds when ${\bf{P}}^{Up}=c{\bf{h}}_{L - 1}$ for some constant $c$.
\end{remark}

\section{Lower Bound to ANC Rate}
We expect to obtain a good AF-based ANC scheme by solving an optimization problem in this section. The SNR function given in (16) is the object function, which is subject to the constraint given in Lemma 1. Note that Lemma 1 only provides a sufficient condition for the power constraint. Therefore, the optimal value of this optimization problem can only be served as a lower bound to the optimal ANC rate.
\begin{equation}
  \left\{
   \begin{aligned}
   &\max &SNR\left( {\left\{ {\beta _k ,k \in {\cal{L}}_1,{\cal{L}}_2 \cdots {\cal{L}}_{L-1} } \right\}} \right)\\
   &s.t.&\beta _k^2  \le \frac{{P_k }}{P_{R,k}+1},k \in {\cal{L}}_1,{\cal{L}}_2 \cdots {\cal{L}}_{L-1}
   \end{aligned}
  \right.
\end{equation}

As shown in [18], a similar optimization problem has a simple solution under the sum power constraint for a two-hop relay network as depicted in Fig. 3. However, the optimization problem given above is essential a fractional programming [22] which is a hard problem. Even for a two-hop network, the above problem is a fractional programming [23] where both the enumerator and denominator are convex functions. There is no such easy approach to solve the problem so far as we know.

\begin{figure}
  \centering
  \includegraphics[width=2.1in,height=1.7in]{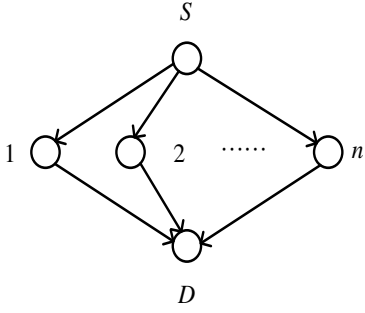}
  \caption{Two-hop parallel relay network with single source-destination pair.}
\end{figure}

Before investigating the lower bound to the optimal ANC rate for the layered relay network with arbitrary layers, we first examine three different AF schemes for two single-source single-destination two-hop parallel relay networks as shown in Fig. 4, and Fig. 6. Then we capture the characterizations of these schemes.
\begin{itemize}
\item Scheme 1: As observed in most conventional amplify-and-forward relaying schemes, all the relay nodes transmit the received signal at the upper bounds of the power constraints. Hence, the amplification gains are
\[
\beta _k  = \beta^{Up} _k,k=1,2\cdots n.
\]
\item Scheme 2: Motivated by the scheme used in layer $l_0$ when we obtain Theorem 1, we employ the pseudo-optimal amplification scheme for this two-hop network, that is
\[
\beta _k  = c\frac{1}{h_{kD} }\sqrt {P_{R,k} }, k=1,2 \cdots n,
\]
where $\displaystyle{c = \min \left\{ \beta^{Up}_k \frac{h_{kD}}{\sqrt{P_{R,k}}}, k=1,2 \cdots n \right\}}$.
\item Scheme 3: The destination node chooses one relay node to transmit the received noisy signal which is referred to as AF with selection. To be specific, only the relay node, with the assistant of which the destination receives the highest SNR value, amplifies and forwards its received noisy signal with the upper bound of the power constraint. Find
\[
k_0= \mathop {\arg }\limits_{k=1,2 \cdots n} \max \left(h_{Sk}h_{kD}\beta^{Up}_k\right)^2,
\]
then choose
\begin{center}
$\beta_{k_0}=\beta^{Up}_{k_0}$ and $\beta_k=0, k \ne k_0$.
\end{center}
\end{itemize}

\begin{figure}
  \centering
  \includegraphics[width=2.1in,height=1.7in]{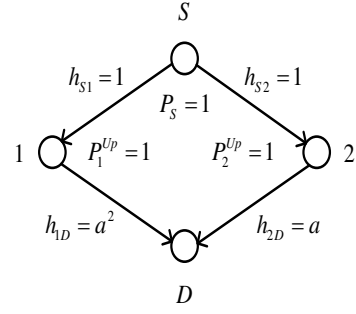}
  \caption{3-layer network with 2 nodes at layer 2.}
\end{figure}

For the network parameters given in Fig. 4, we compute the corresponding amplification gains and the SNR values received at the destination node for the above three schemes. Assume each Gaussian noise introduced by the non-source node is i.i.d. with unit variance. Later, we will explain that the assumption does not lose the generality.
\begin{itemize}
\item Scheme 1:
\begin{center}
$\displaystyle{\beta _1  = \frac{1}{{\sqrt 2 }}}$, $\displaystyle{\beta _2  = \frac{1}{{\sqrt 2 }}}$ and $\displaystyle{SNR = \frac{{\left( {a^2  + a} \right)^2 }}{{a^4  + a^2  + 1}}}$.
\end{center}
\item Scheme 2:

For $0 \le a < 1$,
\begin{center}
  $\displaystyle{c = \frac{{a^2 }}{{\sqrt 2 }}}$, $\displaystyle{\beta _1  = \frac{1}{{\sqrt 2 }}}$, $\displaystyle{\beta _2  = \frac{a}{{\sqrt 2 }}}$, and $\displaystyle{SNR = \frac{{2a^4 }}{{a^4  + 1}}}$,
\end{center}
and for $1 \le a $,
\begin{center}
  $\displaystyle{c = \frac{{a }}{{\sqrt 2 }}}$, $\displaystyle{\beta _1  = \frac{1}{{a \sqrt 2 }}}$, $\displaystyle{\beta _2  = \frac{1}{{\sqrt 2 }}}$, and $\displaystyle{SNR = \frac{{2a^2 }}{{a^2  + 1}}}$.
\end{center}
\item Scheme 3:

For $0 \le a < 1$,
\begin{center}
   $\beta _1  = 0$, $\displaystyle{\beta _2  = \frac{1}{{\sqrt 2 }}}$, and $\displaystyle{SNR = \frac{{a^2 }}{{a^2  + 2}}}$,
\end{center}
and for $1 \le a $,
\begin{center}
  $\displaystyle{\beta_1  = \frac{1}{{\sqrt 2 }}}$, $\displaystyle{\beta _2  = 0}$, and $\displaystyle{SNR = \frac{{a^4 }}{{a^4  + 2}}}$.
\end{center}
\end{itemize}

Then we compute the upper bound of the achievable rate for this network. By Theorem 1,
\begin{center}
$R^{Up}  = \min \left\{ {R^{Up}_1 ,R^{Up}_2 } \right\}$,
\end{center}
where $\displaystyle{R^{Up}_1  = 0.5\log 3}$ and $\displaystyle{R^{Up}_2  = 0.5\log \left( {1 + \left( {a^2  + a} \right)^2 } \right)}$.
The result shows that if $0 \le a < 1.3028$, then $R^{Up}_1$ is greater than $R^{Up}_2$, which means the bottleneck of data transfer is on the second hop, and if $1.3028 \le a$, then $R^{Up}_1$ is no larger than $R^{Up}_2$, which means the bottleneck of data transfer is on the first hop.

In Fig. 5, the numerical result of the achievable rate for each scheme is given. From Fig. 5, we observe that when $0 \le a < 2$, the achievable rates of scheme 1 and 2 are almost the same, while when $2 \le a$, the achievable rate of scheme 2 outperforms that of scheme 1. We are interested in how to select the AF scheme in different cases. Since in low SNR scenario, it is not expected to perform AF as the relaying scheme, to investigate the criterion, we consider the case when $a$ is sufficiently large, i.e., $a \ge 2$. The total noise received at the destination node has two portions, one is the propagation noise from relay nodes denoted by $w_D$, which is referred to as the upper layer noise hereafter, and the other is the noise introduced by the destination node itself which is referred to as the lower layer noise later. We compute the propagation noise power both for scheme 1 and 2. In scheme 1,
\[
E\left[w_D^2\right]=\frac{1}{2}(a^4+a^2),
\]
and in scheme 2,
\[
E\left[w_D^2\right]=a^2.
\]
Since the lower layer noise has unit power, we observe that in both cases, the propagation noises dominate the total noise received at the destination node. Therefore, the AF scheme proposed in our paper is expected to be more effective in dealing with the upper layer noises. Note that the dominance only depends on the ratio of the variances of the upper layer and lower layer noises, hence assuming all the Gaussian noises have unit variance is without loss of generality.

\begin{figure}
  \centering
  \includegraphics[width=3in,height=2.5in]{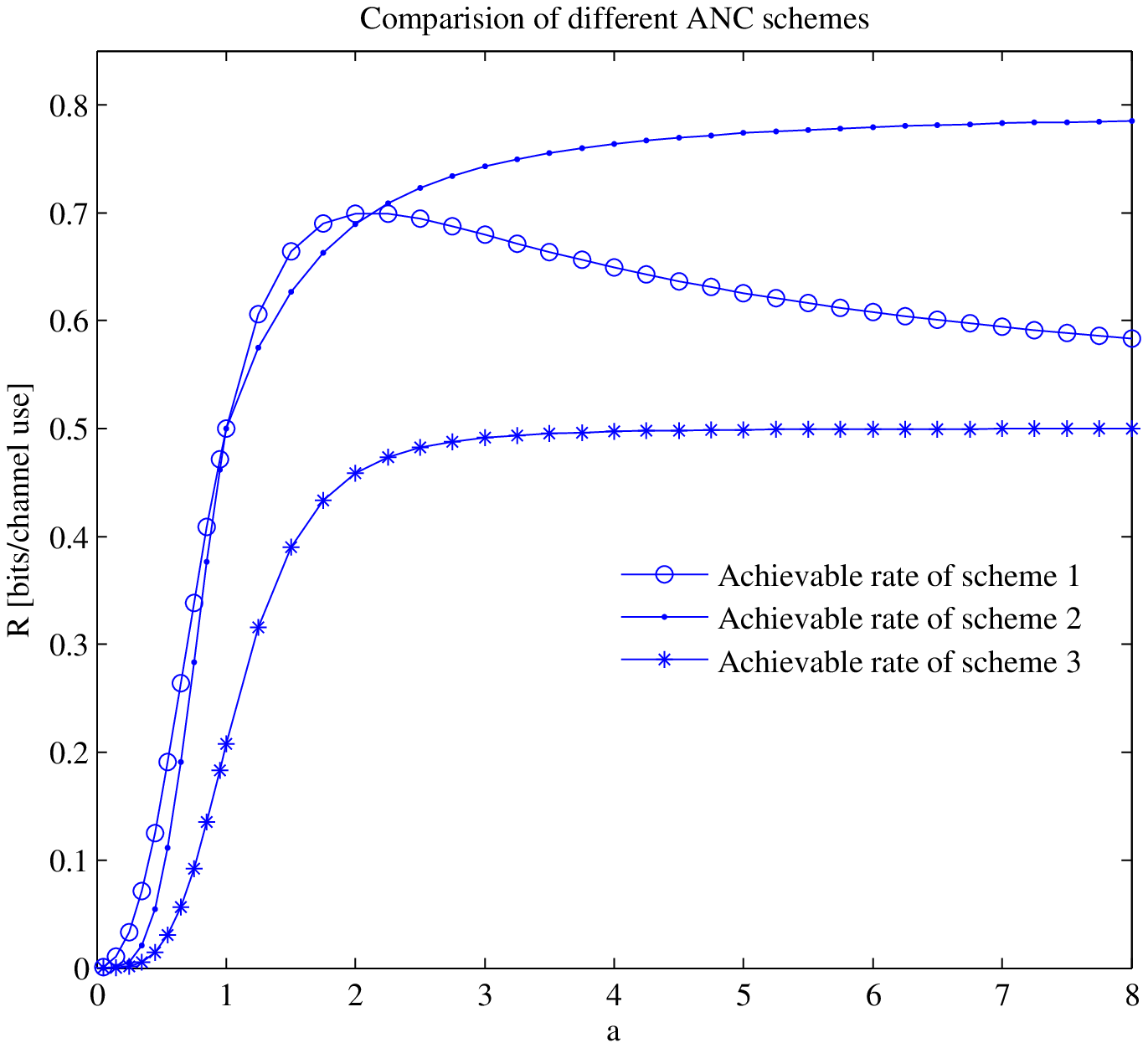}
  \caption{Achievable rates of different AF schemes.}
\end{figure}

\begin{figure}
  \centering
  \includegraphics[width=2.1in,height=1.7in]{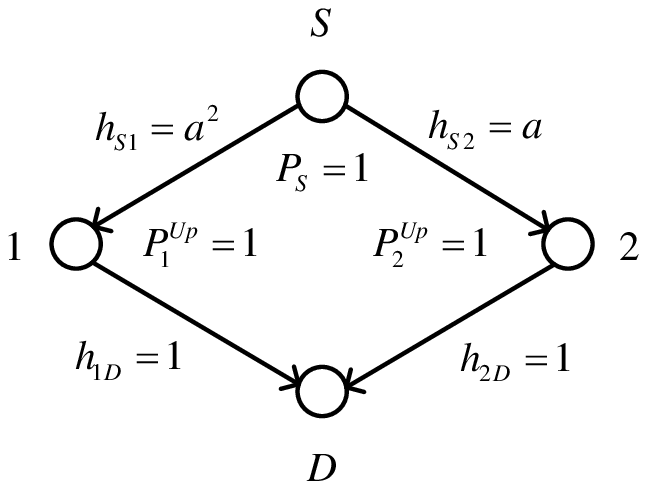}
  \caption{3-layer network with 2 nodes at layer 2.}
\end{figure}

\begin{figure}
  \centering
  \includegraphics[width=3in,height=2.5in]{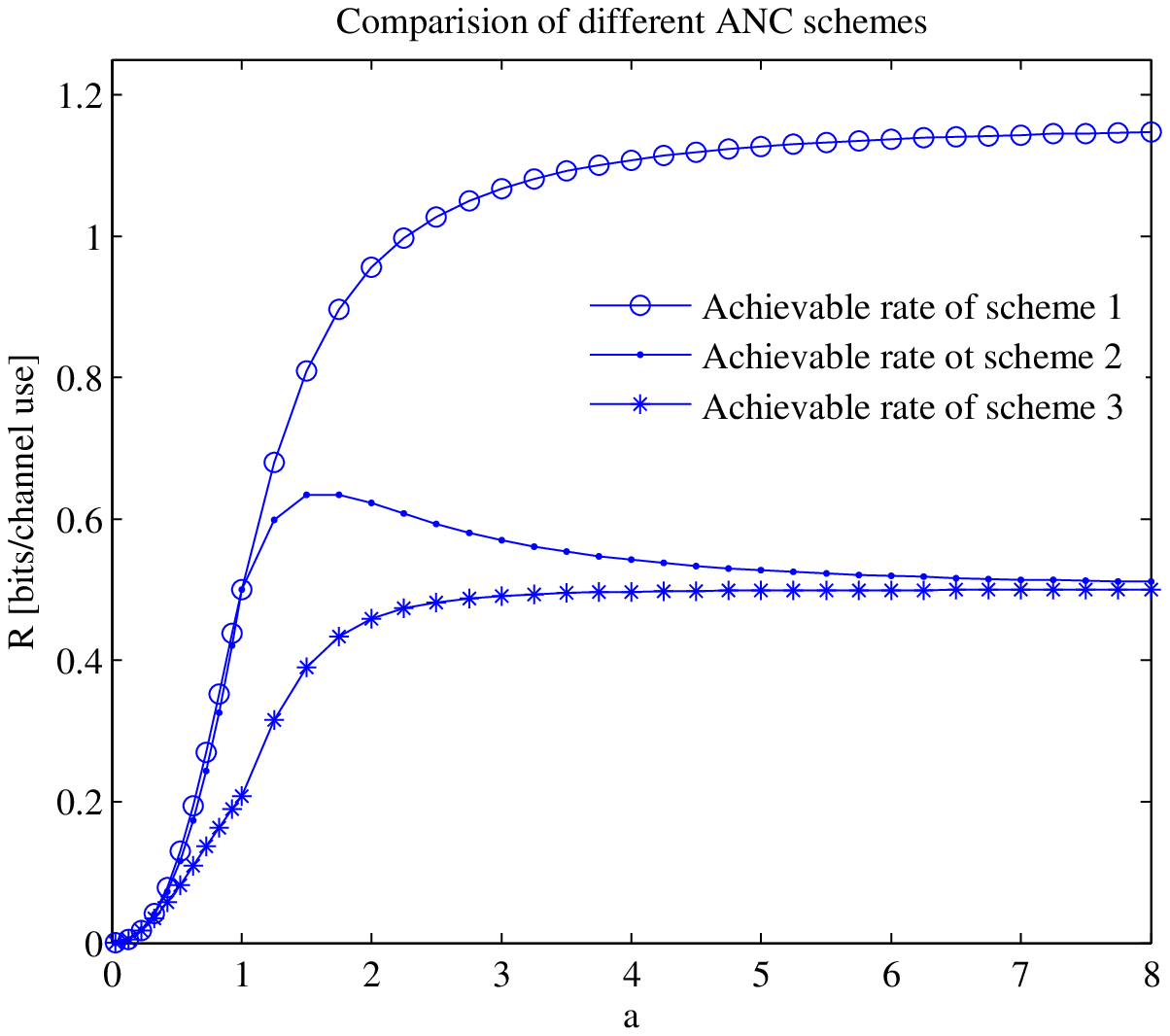}
  \caption{Achievable rates of different AF schemes.}
\end{figure}

For the network parameters given in Fig. 6, we compute the corresponding amplification gains with the above three schemes.
\begin{itemize}
\item Scheme 1:
\begin{center}
$\displaystyle{\beta _1  = \frac{1}{{\sqrt {1+a^4} }}}$, $\displaystyle{\beta _2  = \frac{1}{{\sqrt {1+a^2} }}}$, and $\displaystyle{
SNR = \frac{{\left( {\frac{{a^2 }}{{\sqrt {1 + a^4 } }} + \frac{a}{{\sqrt {1 + a^2 } }}} \right)^2 }}{{\frac{{1 }}{{1 + a^4 }} + \frac{{1 }}{{1 + a^2 }} + 1}}}$.
\end{center}
\item Scheme 2:

For $0 \le a < 1$,
\begin{center}
  $\displaystyle{c = \frac{{1 }}{{a\sqrt{1+a^2} }}}$, $\displaystyle{\beta _1  = \frac{a}{{\sqrt {1+a^2} }}}$, $\displaystyle{\beta _2  = \frac{1}{{\sqrt {1+a^2} }}}$, and $\displaystyle{SNR = \frac{{(a^3+a)^2 }}{{2(a^2  + 1)}}}$,
\end{center}
and for $1 \le a $,
\begin{center}
  $\displaystyle{c = \frac{{1 }}{{a^2\sqrt{1+a^4} }}}$, $\displaystyle{\beta _1  = \frac{1}{{ \sqrt {1+a^4} }}}$, $\displaystyle{\beta _2  = \frac{1}{{a\sqrt {1+a^4} }}}$, and $\displaystyle{SNR = \frac{{(a^3+a)^2 }}{{a^6  +2a^2+ 1}}}$.
\end{center}
\item Scheme 3:

For $0 \le a < 1$,
\begin{center}
   $\beta _1  = 0$, $\displaystyle{\beta _2  = \frac{1}{{\sqrt{1+a^2} }}}$, and $\displaystyle{SNR = \frac{{a^2 }}{{a^2  + 2}}}$,
\end{center}
and for $1 \le a $,
\begin{center}
  $\displaystyle{\beta_1 = \frac{{1 }}{{\sqrt {1+a^4} }}}$, $\displaystyle{\beta _2  = 0}$, and $\displaystyle{SNR = \frac{{a^4 }}{{a^4  + 2}}}$.
\end{center}
\end{itemize}

Then we compute the upper bound of the achievable rate for this network. By Theorem 1,
\begin{center}
$R^{Up}  = \min \left\{ {R^{Up}_1 ,R^{Up}_2 } \right\}$,
\end{center}
where $\displaystyle{R^{Up}_1  = 0.5\log \left( {1 + a^2  + a^4 } \right)}$ and $\displaystyle{R^{Up}_2  = 0.5\log 5}$.
The result shows that if $0 \le a < 1.2496$, then $R^{Up}_1$ is less than $ R^{Up}_2$, which means the bottleneck of data transfer is on the first hop, and if $1.2496 \le a$, then $R^{Up}_1$ is no less than $ R^{Up}_2$, which means the bottleneck of data transfer is on the second hop.

In Fig. 7, the numerical result of the achievable rate for each scheme is given. From Fig. 7, we observe that when $0 \le a < 1$, the achievable rates of scheme 1 and 2 are almost the same while when $1 \le a$, the achievable rate of scheme 1 outperforms that of scheme 2. We are still interested in how to select the AF scheme in different cases. As explained previously, we consider the case when $a$ is sufficiently large, i.e., $a \ge 1$. We compute the propagation noise power both for scheme 1 and 2. In scheme 1,
\[
E\left[w_D^2\right]={\frac{{1}}{{1 + a^4 }} + \frac{{1 }}{{1 + a^2 }}},
\]
and in scheme 2,
\[
E\left[w_D^2\right]=\frac{{1}}{{1 + a^4 }}+\frac{{1}}{{a^2(1 + a^4) }}.
\]
We observe that in both cases, the lower layer noise dominates the total noise. Therefore, from the numerical result, the scheme 1 is expected to be more effective in dealing with the lower layer noise.

Now let us summarize the results drawn from the above analysis. The first observation is that the more relay nodes assist the transmission the better performance we can obtain. The reason is that the source signals retransmitted by the relay nodes are all related; however, the noise signals are independent of each other. Therefore, from the aspect of cooperation communications, the more copies of the original signals received at the destination node the lager diversity gain will be obtained. Second, if the relay nodes always amplify and forward the received signal to the maximum possible value, then it may result in suboptimal end-to-end throughput. This observation leads to an interesting problem that under what conditions the two schemes will have a better performance? From the above two examples, we conjecture that scheme 1 is more effective in managing lower layer noises and inversely, scheme 2 is more valid in dealing with upper layer noises. Finally, the first two AF schemes mentioned above can constitute a mixed ANC scheme which can be applied to a layered relay network with arbitrary layers. The relay nodes at each layer may select either of the conventional scheme (scheme 1) or the new scheme (scheme 2) proposed in our paper. However, in a layered relay network, the selection of the scheme for one layer also depends on the schemes applied by the other layers. Therefore, it is complicated to provide a localized criterion to select which scheme at each layer.

Let us consider the following AF-based ANC scheme motivated by the optimal solution in the ideal network analyzed in Theorem 1. With such ANC scheme, a lower bound to the optimal ANC rate is derived. To simplify the computation of the lower bound, we first define
\begin{equation}
\delta_0=
\mathop {\max }\limits_{j \in {\cal{L}}_1, \cdots , {\cal{L}}_{l_0 - 1},{\cal{L}}_{l_0 + 1}, \cdots ,{\cal{L}}_L} \delta_j.
\end{equation}
The amplification gain at node $k,k \in {\cal{L}}_1,{\cal{L}}_2 \cdots {\cal{L}}_{l_0 - 1}$ is chosen as
\begin{equation}
\beta _k  = \sqrt {\frac{{P_k^{Up} }}{{\left( {1 + \delta _0 } \right)P_{R,k} }}} \le \sqrt {\frac{{P_k^{Up} }}{{\left( {1 + \delta _k } \right)P_{R,k} }}} = \beta^{Up}_k,
\end{equation}
and at node $k,k\in{\cal{L}}_{l_0 + 1} \cdots {\cal{L}}_{L - 1}$ is chosen as some positive constant such that
\begin{equation}
\beta _k  = const. \le \beta^{Up}_k.
\end{equation}
Note that the amplification vector selected at layer $l_0$ depends on these factors and moreover, no matter what feasible gains are selected will not affect the resulting upper bound.
The amplification vector at layer $l_0$ is chosen as
\begin{equation}
{\bm{\beta }}_{l_0}  = c_1 {\bf{G}}^{ - 1} {\bf{P}}_{R,l_0},
\end{equation}
where ${\bm{\beta }}_{l_0}$ and ${\bf{G}}$ are defined in the proof of Theorem $1$, and $c_1$ can be determined as
\begin{equation}
c_1  = \min \left\{\beta^{Up}_k {\frac{{g_k }}{\sqrt{P_{R,k} }}},k\in {\cal{L}}_{l_0} \right\}.
\end{equation}

Note that, in the above ANC scheme, both the conventional and the pseudo-optimal amplification scheme are adopted. Although it is not expected to obtain the optimal ANC rate through a fixed scheme, the corresponding achievable rate of such ANC scheme can be served as a lower bound.

\newtheorem{thm2}{Theorem}
\begin{thm}[Lower Bound to ANC Rate]
A lower bound of the ANC rate of a Gaussian layered relay network with $L$ layers under the transmitting power constraints is as follows.
\begin{align}
R&^{Low} = \nonumber \\
& \mathop {\max }\limits_{l_0  = 1,2 \cdots L} \frac{1}{2}\log \left( {1 + \frac{{\frac{{{\bf{P}}_{R,l_0 }^T {\bf{P}}_{R,l_0 } }}{{\left( {1 + \delta_0 } \right)^{l_0  - 1} }}}}{{\left[ {1 - \frac{1}{{\left( {1 + \delta_0 } \right)^{l_0  - 1} }}} \right]{\bf{P}}_{R,l_0 }^T {\bf{P}}_{R,l_0 }  + c_3 }}} \right),\nonumber
\end{align}
where $c_2 =
E\left[ {\left( {\sum\limits_{l = l_0  + 1}^{L - 1} {{\bf{h}}_{L - 1}^T {\bf{B}}_{L - 1}  \cdots {\bf{H}}_l {\bf{B}}_l {\bf{z}}_l }  } \right)^2 } \right]
$ is a bounded constant and $c_3  = 1 + \displaystyle\frac{{c_2 +1}}{{c_1 ^2 {\bf{P}}_{R,l_0}^T {\bf{P}}_{R,l_0} }}
$.
\end{thm}
\begin{IEEEproof}
The proof is given in the appendix C.
\end{IEEEproof}

Note that $c_2$ is a bounded positive constant depending on the selection of amplification gains in nodes at layer $l_0+1$ to $L-1$. Finally, we should emphasize that the lower bound obtained in the above theorem is a universal one since it is not required that $\delta_0$ should be a small positive value.

\section{Applications of AF-based ANC scheme}
To the best of our knowledge, AF relay design for a layered relay network has been the focus of mush recent research. Thus by taking pseudo-optimal amplification scheme at certain layers,
we have an alternative multi-hop ANC scheme as opposite to the scheme with all relay nodes amplifying the received signals to the
maximum [15]. A problem is whether pseudo-optimal amplification scheme can make things better than conventional AF scheme. We turn to
two particular classes of layered relay networks and compare the resulting lower bound of AF rates to our upper bound.

\subsection{Application to Two-Hop Network}
We extend the results derived from the previous examples to a two-hop parallel relay network with $n$ relay nodes.
Note that scheme 2 can be viewed as a pseudo-optimal amplification scheme for such network. Denote by ${\bf{B}}^{(1)}$ and ${\bf{B}}^{(2)}$ the two schemes and by $w_D^{(1)}$ and $w_D^{(2)}$ the upper layer noises corresponding to these two schemes respectively. Hence, we have
\begin{equation}
E\left[(w^{(i)}_D)^2\right]={\bf{h}}_1^T {\bf{B}}^{\left( i \right)} {\bf{B}}^{\left( i \right)}{\bf{h}}_1,{\rm{    }}i=1,2,
\end{equation} and the lower layer noise introduced by the destination node itself has, as assumed, a unit power.
\begin{equation}
E\left[z_D^2\right]=1.
\end{equation}
It is easy to verify that $E\left[(w^{(1)}_D)^2\right]$ is no less than $ E\left[(w^{(2)}_D)^2\right]$.

\newtheorem{definition4}{Definition}
\begin{definition}
We call the upper layer noise dominates the total noises received at the destination in a two-hop relay network with AF relays when \[
E\left[(w^{(1)}_D)^2\right] \ge E\left[(w^{(2)}_D)^2\right]\ge 1
.\] Reversely, we call the lower layer noise dominates the total noises when
\[E\left[(w^{(2)}_D)^2\right] \le E\left[(w^{(1)}_D)^2\right]\le1.\]
\end{definition}

As we mentioned before, the individual power constraints make the problem of deriving the exact optimal AF rate complicated to be handled.  However, we still obtain some interesting results on the performance of achievable AF rate shown as follows.

\newtheorem{thm3}{Theorem}
\begin{thm}
For a two-hop relay network with AF relays under individual power constraints, there exist AF schemes accodrding to different senarios
when either the upper layer noise or the lower layer noise dominates the total noises recived at the destination node such that the gaps between the
the corresponding achievable rates $R$'s and the optimal AF rate $R_{opt}$ are at most $1/2$ bit.
\end{thm}
\begin{IEEEproof}
For a two-hop relay network, we see that the achievable rate can be easily derived without employing the lower bound obtained in Theorem 2. For the first scenario, we compute the achievable rate with the pseudo-optimal amplification scheme, i.e., ${\bf{B}}^{(2)}$.
\begin{align}
 R &= \frac{1}{2}\log \left( {1 + \frac{{{\bf{h}}_1^T {\bf{B}}^{\left( 2 \right)} {\bf{h}}_0 {\bf{h}}_0^T {\bf{B}}^{\left( 2 \right)} {\bf{h}}_1 }P_{S}}{{E\left[ {(w_D^{(2)} )^2 } \right] + 1}}} \right) \nonumber\\
  &= \frac{1}{2}\log \left( {1 + \frac{{{{{\bf{h}}_1^T {\bf{B}}^{\left( 2 \right)} {\bf{h}}_0 {\bf{h}}_0^T {\bf{B}}^{\left( 2 \right)} {\bf{h}}_1 } \mathord{\left/
 {\vphantom {{{\bf{h}}_1^T {\bf{B}}^{\left( 2 \right)} {\bf{h}}_0 {\bf{h}}_0^T {\bf{B}}^{\left( 2 \right)} {\bf{h}}_1 } {E\left[ {(w_D^{(2)} )^2 } \right]}}} \right.
 \kern-\nulldelimiterspace} {E\left[ {(w_D^{(2)} )^2 } \right]}}}}{{1 + {1 \mathord{\left/
 {\vphantom {1 {E\left[ {(w_D^{(2)} )^2 } \right]}}} \right.
 \kern-\nulldelimiterspace} {E\left[ {(w_D^{(2)} )^2 } \right]}}}}} P_{S}\right) \nonumber
\end{align}
\begin{align}
  &\mathop  \ge \limits^{\left( a \right)} \frac{1}{2}\log \left( {1 + \frac{{{{{\bf{h}}_1^T {\bf{B}}^{\left( 2 \right)} {\bf{h}}_0 {\bf{h}}_0^T {\bf{B}}^{\left( 2 \right)} {\bf{h}}_1 } \mathord{\left/
 {\vphantom {{{\bf{h}}_1^T {\bf{B}}^{\left( 2 \right)} {\bf{h}}_0 {\bf{h}}_0^T {\bf{B}}^{\left( 2 \right)} {\bf{h}}_1 } {E\left[ {(w_D^{(2)} )^2 } \right]}}} \right.
 \kern-\nulldelimiterspace} {E\left[ {(w_D^{(2)} )^2 } \right]}}}}{2}} P_{S}\right) \nonumber\\
  &\ge \frac{1}{2}\log \left( {1 + \frac{{{\bf{h}}_1^T {\bf{B}}^{\left( 2 \right)} {\bf{h}}_0 {\bf{h}}_0^T {\bf{B}}^{\left( 2 \right)} {\bf{h}}_1 }P_{S}}{{E\left[ {(w_D^{(2)} )^2 } \right]}}} \right) - \frac{1}{2} \nonumber\\
 &\mathop  \ge \limits^{\left( b \right)} R_{opt}  - \frac{1}{2}
\end{align}
where (a) follows from the assumption that $E\left[(w^{(2)}_D)^2\right] \ge 1$ and (b) follows from Theorem 1 that
\[
R_{opt} \le R^{Up}_1 = \frac{1}{2}\log \left( {1 + \frac{{{\bf{h}}_1^T {\bf{B}}^{\left( 2 \right)} {\bf{h}}_0 {\bf{h}}_0^T {\bf{B}}^{\left( 2 \right)} {\bf{h}}_1 }P_{S}}{{E\left[ {(w_D^{(2)} )^2 } \right]}}} \right).
\]

For the second scenario, we compute the achievable rate with the conventional scheme, i.e., ${\bf{B}}^{(1)}$.

\begin{align}
 R &= \frac{1}{2}\log \left( {1 + \frac{{{\bf{h}}_1^T {\bf{B}}^{\left( 1 \right)} {\bf{h}}_0 {\bf{h}}_0^T {\bf{B}}^{\left( 1 \right)} {\bf{h}}_1 }P_{S}}{{E\left[ {(w_D^{(1)} )^2 } \right] + 1}}} \right)\nonumber \\
  &\mathop  \ge \limits^{\left( a \right)}\frac{1}{2}\log \left( {1 + \frac{{{\bf{h}}_1^T {\bf{B}}^{\left( 1 \right)} {\bf{h}}_0 {\bf{h}}_0^T {\bf{B}}^{\left( 1 \right)} {\bf{h}}_1 }P_{S}}{2}} \right) \nonumber\\
  &\ge \frac{1}{2}\log \left( {1 + {\bf{h}}_1^T {\bf{B}}^{\left( 1 \right)} {\bf{h}}_0 {\bf{h}}_0^T {\bf{B}}^{\left( 1 \right)} {\bf{h}}_1 } P_{S}\right) - \frac{1}{2} \nonumber\\
  &\mathop  \ge \limits^{\left( b \right)} R_{opt}  - \frac{1}{2}
\end{align}
where (a) follows from the assumption that $E\left[(w^{(1)}_D)^2\right]\le1$ and (b) follows from Theorem 1 that
\[
R_{opt} \le R^{Up}_2 = \frac{1}{2}\log \left( {1 + {\bf{h}}_1^T {\bf{B}}^{\left( 1 \right)} {\bf{h}}_0 {\bf{h}}_0^T {\bf{B}}^{\left( 1 \right)} {\bf{h}}_1 } P_{S}\right).
\]

Then we complete the proof.
\end{IEEEproof}

Note that Theorem 3 rigorously interpret the intuition obtained from the examples in section IV. We are also interested in the case beyond these two scenarios, i.e., when $E\left[(w^{(1)}_D)^2\right]\ge 1 \ge E\left[(w^{(2)}_D)^2\right] $. Are the performances of the two schemes in such general scenario also good enough? We draw the conclusion in the following theorem.
\newtheorem{thm4}{Theorem}
\begin{thm}
For a two-hop relay network with AF relays under individual power constraints, with either the pseudo-optimal amplification scheme or the conventional one,
the gaps between the corresponding achievable rates $R$'s and the optimal AF rate $R_{opt}$ are upper bounded by a constant when neither the upper layer noise nor the lower layer noise dominates the total
received noises at the destination.
\end{thm}
\begin{IEEEproof}
To verify the statement, we compare achievable rates corresponding to scheme $1$ and $2$ with two upper bounds $R^{Up}_2$ and $R^{Up}_1$ respectively. The achievable rate with the conventional scheme ${\bf{B}}^{(1)}$ is as follows.
\begin{align}
 R &= \frac{1}{2}\log \left( {1 + \frac{{{\bf{h}}_1^T {\bf{B}}^{\left( 1 \right)} {\bf{h}}_0 {\bf{h}}_0^T {\bf{B}}^{\left( 1 \right)} {\bf{h}}_1 }P_{S}}{{E\left[ {(w_D^{(1)} )^2 } \right] + 1}}} \right) \nonumber\\
 &\mathop  \ge \limits^{\left( a \right)} \frac{1}{2}\log \left( {1 + \frac{{{\bf{h}}_1^T {\bf{B}}^{\left( 1 \right)} {\bf{h}}_0 {\bf{h}}_0^T {\bf{B}}^{\left( 1 \right)} {\bf{h}}_1 }P_{S}}{{{{E\left[ {(w_D^{(1)} )^2 } \right]} \mathord{\left/
 {\vphantom {{E\left[ {(w_D^{(1)} )^2 } \right]} {E\left[ {(w_D^{(2)} )^2 } \right]}}} \right.
 \kern-\nulldelimiterspace} {E\left[ {(w_D^{(2)} )^2 } \right]}} + 1}}} \right) \nonumber
\end{align}
\begin{align}
&\mathop  \ge \limits^{\left( b \right)} \frac{1}{2}\log \left( {1 + \frac{{{\bf{h}}_1^T {\bf{B}}^{\left( 1 \right)} {\bf{h}}_0 {\bf{h}}_0^T {\bf{B}}^{\left( 1 \right)} {\bf{h}}_1 }P_{S}}{{\varepsilon + 1}}} \right) \nonumber\\
& \ge \frac{1}{2}\log \left( {1 + {\bf{h}}_1^T {\bf{B}}^{\left( 1 \right)} {\bf{h}}_0 {\bf{h}}_0^T {\bf{B}}^{\left( 1 \right)} {\bf{h}}_1 } P_{S}\right) - \frac{1}{2}\log \left( {1 + \varepsilon} \right) \nonumber\\
 &\mathop  \ge \limits^{\left( c \right)} R_{opt}  - \frac{1}{2}\log \left( {1 + \varepsilon} \right),
\end{align}
where $\displaystyle{\varepsilon= \max \left\{\left(\frac{\beta^{(1)}_k}{\beta^{(2)}_k}\right)^2,k=1,2\cdots n\right\}}$, (a) follows from the assumption $1 \ge E\left[(w^{(2)}_D)^2\right] $, (b) follows from
\begin{equation}
\frac{{E\left[ {(w_D^{(1)} )^2 } \right]}}{{E\left[ {(w_D^{(2)} )^2 } \right]}} = \frac{{{\bf{h}}_1^T {\bf{B}}^{\left( 1 \right)} {\bf{B}}^{\left( 1 \right)} {\bf{h}}_1 }}{{{\bf{h}}_1^T {\bf{B}}^{\left( 2 \right)} {\bf{B}}^{\left( 2 \right)} {\bf{h}}_1 }} \le \varepsilon,
\end{equation}
where the last inequality results from the generalized Rayleigh quotient [27], and (c) follows from the same argument as (b) in (40). Similarly, the achievable rate with the pseudo-optimal amplification scheme ${\bf{B}}^{(2)}$ is as follows.
\begin{align}
 R& = \frac{1}{2}\log \left( {1 + \frac{{{\bf{h}}_1^T {\bf{B}}^{\left( 2 \right)} {\bf{h}}_0 {\bf{h}}_0^T {\bf{B}}^{\left( 2 \right)} {\bf{h}}_1 }P_{S}}{{E\left[ {(w_D^{(2)} )^2 } \right] + 1}}} \right) \nonumber\\
 &\mathop  \ge \limits^{\left( a \right)} \frac{1}{2}\log \left( {1 + \frac{{{{{\bf{h}}_1^T {\bf{B}}^{\left( 2 \right)} {\bf{h}}_0 {\bf{h}}_0^T {\bf{B}}^{\left( 2 \right)} {\bf{h}}_1 } \mathord{\left/
 {\vphantom {{{\bf{h}}_1^T {\bf{B}}^{\left( 2 \right)} {\bf{h}}_0 {\bf{h}}_0^T {\bf{B}}^{\left( 2 \right)} {\bf{h}}_1 } {E\left[ {(w_D^{(2)} )^2 } \right]}}} \right.
 \kern-\nulldelimiterspace} {E\left[ {(w_D^{(2)} )^2 } \right]}}}}{{1 + {{E\left[ {(w_D^{(1)} )^2 } \right]} \mathord{\left/
 {\vphantom {{E\left[ {(w_D^{(1)} )^2 } \right]} {E\left[ {(w_D^{(2)} )^2 } \right]}}} \right.
 \kern-\nulldelimiterspace} {E\left[ {(w_D^{(2)} )^2 } \right]}}}}} P_{S}\right) \nonumber\\
 &\mathop  \ge \limits^{\left( b \right)} \frac{1}{2}\log \left( {1 + \frac{{{{{\bf{h}}_1^T {\bf{B}}^{\left( 2 \right)} {\bf{h}}_0 {\bf{h}}_0^T {\bf{B}}^{\left( 2 \right)} {\bf{h}}_1 } \mathord{\left/
 {\vphantom {{{\bf{h}}_1^T {\bf{B}}^{\left( 2 \right)} {\bf{h}}_0 {\bf{h}}_0^T {\bf{B}}^{\left( 2 \right)} {\bf{h}}_1 } {E\left[ {(w_D^{(2)} )^2 } \right]}}} \right.
 \kern-\nulldelimiterspace} {E\left[ {(w_D^{(2)} )^2 } \right]}}}}{{1 + \varepsilon  }}} P_{S}\right) \nonumber\\
  &\ge \frac{1}{2}\log \left( {1 + \frac{{{\bf{h}}_1^T {\bf{B}}^{\left( 2 \right)} {\bf{h}}_0 {\bf{h}}_0^T {\bf{B}}^{\left( 2 \right)} {\bf{h}}_1 }P_{S}}{{E\left[ {(w_D^{(2)} )^2 } \right]}}} \right) - \frac{1}{2}\log \left( {1 + \varepsilon } \right) \nonumber\\
  &\mathop  \ge \limits^{\left( c \right)}R_{opt}  - \frac{1}{2}\log \left( {1 + \varepsilon } \right) ,
\end{align}
where (a) follows from the assumption $E\left[(w^{(1)}_D)^2\right]\ge 1 $, (b) follows from (42) and (c) follows from the same argument as (b) in (39).

Therefore, we complete the proof.
\end{IEEEproof}

\subsection{Application to Generalized High-SNR Regime}
In this subsection, a special SNR regime is proposed, under which the ANC scheme proposed in section IV is shown to have a good performance. The scenario considered in this paper can be viewed as an extension of that discussed in [15]. In the similar manner as in [15], we define the generalized high-SNR regime as follows.
\begin{definition}[Generalized High-SNR Regime]
 We call the layered relay network is in the generalized high-SNR regime if
\begin{equation}
 \delta_0 \le \delta ,
\end{equation}
for some small positive $
\delta$, where $\delta_0$ is defined in (36).
\end{definition}

From the definition, it is assumed that the received SNR at each node $k$, $k \notin {\cal{L}}_{l_0}$ is large which is referred to as the high-SNR condition in the sequel. Therefore, the above SNR regime is an extension of the high-SNR regime defined in [15], where only the destination node does not satisfy the high-SNR condition. If we set $l_0$ equal to $L$, the generalized high-SNR regime is degraded to the high-SNR regime. From the perspective of the optimization problem, we will show that the AF-based ANC scheme proposed previously can be utilized as a good suboptimal solution for (31). It is expected that the achievable rate of such ANC scheme may asymptotically approach the upper bound of the ANC rate within a small gap and thus close to the optimal ANC rate.

Two cases when the generalized high-SNR regime can be satisfied are considered. If assume except the nodes on layer $l_0-1$, all the other nodes have sufficiently large transmitting powers and the transmitting powers of these nodes tend to infinity at the same rate as $\delta  \to 0$, the network is in the generalized high-SNR regime. Under these assumptions, we have $P_{R,k}  = const. , k \in {\cal{L}}_{l_0}$, $c_2=const.$, and $c_1 \to \infty$ as $\delta \to 0$. It is not complicated to verify that
\begin{align}
 \mathop {\lim }\limits_{\delta  \to 0} \Delta & = \mathop {\lim }\limits_{\delta  \to 0} R^{Up}  - R^{Low} \nonumber \\
 & = \mathop {\lim }\limits_{\delta  \to 0} \frac{1}{2}\log \left( {\frac{{1 + {\bf{P}}_{R,l_0 }^T {\bf{P}}_{R,l_0 } }}{1 + \frac{{{\frac{{{\bf{P}}_{R,l_0 }^T {\bf{P}}_{R,l_0 } }}{{\left( {1 + \delta } \right)^{l_0  - 1} }}}}}{{{\left[ {1 - \frac{1}{{\left( {1 + \delta } \right)^{l_0  - 1} }}} \right]{\bf{P}}_{R,l_0 }^T {\bf{P}}_{R,l_0 }  + c_3 }}}}} \right)\nonumber \\
  &= 0
\end{align}

The result implies that the achievable ANC rate asymptomatically approaches the upper bound. This analytical result can be justified intuitively as follows. The total noise received at the destination node is separated into three parts. We first make a comment on the part consisting of the noises introduced by the nodes on layer $1$ to layer $l_0-1$. As the source transmits the signal with large power, (if $l_0 \ne 1$) and all the introduced Gaussian noises have finite powers, this part of noises seems to be small compared with the signal received at nodes $\in {\cal{L}}_{l_0}$ and also at the destination node. Then we discuss the second part consisting of the noises introduced by the nodes at layer $l_0$. As the generalized high-SNR regime indicated, the nodes on layer $l_0-1$ have limited transmitting powers, each node on layer $l_0$ receives a noisy signal with finite powers. Furthermore, as mentioned before, the propagation noise is only a small fraction of it. Meanwhile, the noises introduced at layer $l_0$ have powers on the same level as the received signal and hence this part of noises cannot be ignored at the destination node. Finally, we consider the last part of noises introduced by the nodes at layer $l_0+1$ to layer $L$. Since nodes at layer $l_0$ possess large transmitting powers, the case is similar as the first part. Hence the introduced noises of these nodes are small compared with the transmitted signal from layer $l_0$, which implies that the propagated noises from layer $l_0$ dominate the total noise received at the destination. Therefore, the corresponding achievable ANC rate may asymptotically approach the upper bound obtained from the ideal network under the assumption that only the nodes on layer $l_0$ introduce the noises.

Next we discuss another case. Assume all the nodes both on layer $l_0-1$ and $l_0$ have finite transmitting powers, while any other nodes have sufficiently large transmitting powers. But we assume the number of nodes at the $l_0$th layer is sufficiently large which is similar to the assumption in [7]. Hence, there are indeed two modes the nodes satisfying the high-SNR condition. First, the nodes on layer $l, l=1 \cdots l_0-1,l_0+2\cdots L$ satisfy the high-SNR condition as the previous case. Second, the received signal powers at the nodes on layer $l_0+1$ increases with the number of the nodes on layer $l_0$. Consequently, the generalized high-SNR regime can be satisfied.

To emphasize on the difference between these two modes, let
\begin{equation}
\mathop {\min }\limits_{j \in {\cal{L}}_1 \cdots {\cal{L}}_{l_0-1},{\cal{L}}_{l_0+2} \cdots {\cal{L}}_{L } } P_{R,j}  \ge \frac{1}{{\delta '}},
\end{equation}
for some small $\delta ' > 0$. Then we have,
\begin{align}
&\mathop {\lim }\limits_{n_{l_0}  \to \infty }\mathop {\lim }\limits_{\delta ' \to 0} \Delta \nonumber\\
&= \displaystyle{\mathop {\lim }\limits_{n_{l_0}  \to \infty } \mathop {\lim }\limits_{\delta ' \to 0} \frac{1}{2}\log \left( {\frac{{1 + {\bf{P}}_{R,l_0}^T {\bf{P}}_{R,l_0} }}{{1 + \frac{{{\bf{P}}_{R,l_0}^T {\bf{P}}_{R,l_0} }}{{\left[ {\left( {1 + \delta '} \right)^{l_0-1}  - 1} \right]{\bf{P}}_{R,l_0}^T {\bf{P}}_{R,l_0}  + c_5 }}}}} \right)}\nonumber\\
& = \mathop {\lim }\limits_{n_{l_0}  \to \infty } \frac{1}{2}\log \left( {\frac{{1 + {\bf{P}}_{R,l_0}^T {\bf{P}}_{R,l_0} }}{{1 + \frac{\displaystyle{{{\bf{P}}_{R,l_0}^T {\bf{P}}_{R,l_0} }}}{\displaystyle{{{1 + \frac{c_2+1}{{c_1 ^2 {\bf{P}}_{R,l_0}^T {\bf{P}}_{R,l_0} }}}}}}}}} \right)\nonumber\\
&=0
\end{align}
where $c_5  = \left( {1 + \frac{\displaystyle{c_2+1}}{\displaystyle{{c_1 ^2 {\bf{P}}_{R,l_0}^T {\bf{P}}_{R,l_0} }}}} \right)\left( {1 + \delta '} \right)^{l_0-1}$.

The result also implies that the achievable ANC rate asymptomatically approaches the upper bound. Furthermore, we observe that with the number of the nodes in layer $l_0$ increasing, the upper bound of ANC rate also increases and thus the optimal ANC rate.

Finally, two examples are utilized to illustrate the performance of the ANC scheme in a $4$-layer network for the generalized high-SNR regime with $l_0=2$. The numeric results show that in both cases considered, the achievable ANC rate can asymptotically achieve the upper bound of the ANC rate.

\begin{figure}
  \centering
  \includegraphics[width=2in,height=1.5in]{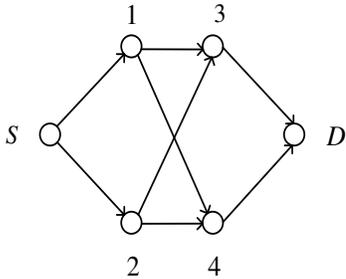}
  \caption{4-layer network with 2 nodes at layer 2.}
\end{figure}

\noindent{\textbf{Example 1}: 4-Layer Network with 2 relays at layer 2}

A 4-layer relay network with $2$ relay nodes on layer $2$ is depicted in Fig. 8. To satisfy the generalized high-SNR regime, assume relay nodes $3$ an $4$ possess sufficiently large transmitting powers. The lower bound of the achievable ANC rate is expected to increase with the transmitting powers of the source node.

\begin{figure}
  \includegraphics[width=3.5in,height=3in]{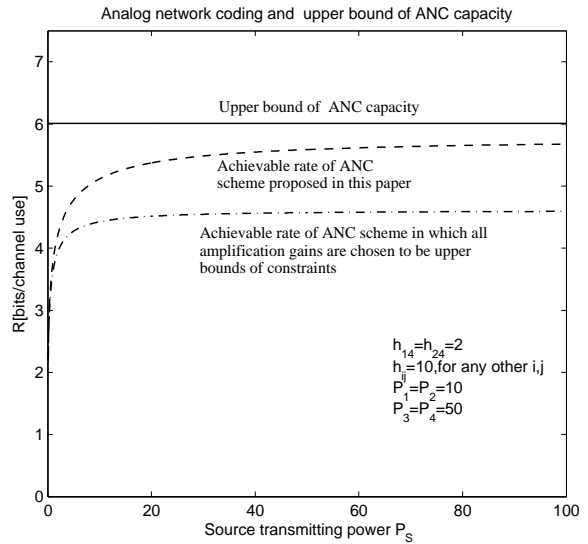}
  \caption{ Achievable ANC rate and upper bound of ANC capacity in 4-layer network.}
\end{figure}

From Fig. 9 we observe the following:

1)	For $P_{R,3}$ and $P_{R,4}$ constants, the achievable ANC rate asymptotically approaches the upper bound of ANC rate. The achievable ANC
rate approaches the capacity to within one bit as $P_S$ larger than $10$, and is within a small fraction of a bit for $P_S $ greater than $100$.

2)	Fig. 9 also shows the achievable ANC rate by setting the amplification gains to the upper bounds of the power
constraints given in Lemma $1$. There is a constant gap approximate $1.5$ bits from the upper bound.

\begin{figure}
  \centering
  \includegraphics[width=2in,height=1.5in]{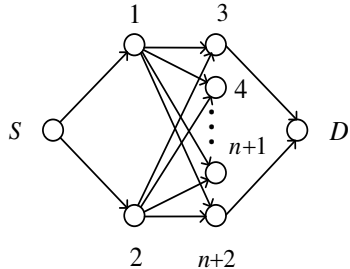}
  \caption{4-layer network with n nodes at layer 2.}
\end{figure}

\noindent{\textbf{Example 2}: 4-Layer Network with n relays at layer 2}

We next present the performance of ANC in the 4-layer network with $n$ relays on layer $2$ depicted in Fig. 10. As the number of the relay nodes $n$ sufficiently large, the generalized high-SNR regime can be satisfied.

The result shown in Fig. 11 implies that the gap between the achievable ANC rate and the upper bound
decreases within one bit as $n$ lager than $5$, and within a small fraction of a bit for $n$ greater than $40$ as the transmitting powers of the nodes on layer 2 are upper bounded by 2.

\begin{figure}
  \includegraphics[width=3.5in,height=3in]{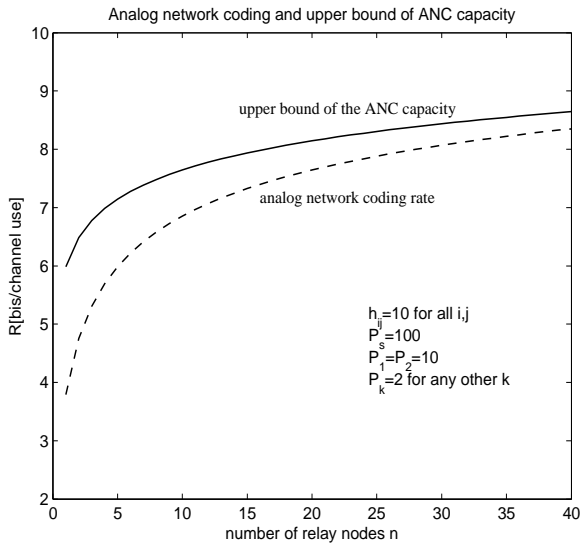}
  \caption{Upper bound of ANC capacity and increase of achievable ANC rate with number of relays at layer 2 in 4-layer
  network.}
\end{figure}


\section{Extension to Two-hop MAC Network}
In this section, we progressively study the performance of the AF scheme in a two-hop parallel MAC depicted in Fig. 12. The structure of the optimal rate region of this network is first characterized by an asymptotical approach. Then a practical low complexity AF-based ANC scheme is proposed.
\subsection{Achievable Rate Region of Two-Hop MAC via AF}
For simplicity, we focus on a two-source case. With the help of the relay nodes, the two source nodes communicate with the destination node individually. Assume each non-source node introduces an i.i.d. Gaussian noise drawn according to the distribution ${\cal{N}}(0,1)$. With the superposition property of the wireless channel, the received signals at each relay node is the linear combination of the signals from the two source nodes and the Gaussian noise. To perform ANC, the relay nodes adopt the AF relay scheme. Hence the received signal at the destination node is given as follows.
\begin{equation}
y_D = {\bf{h}}_1^T {\bf{B}} {\bf{h}}_{01} x_{S_1 }  + {\bf{h}}_1^T {\bf{B}} {\bf{h}}_{02} x_{S_2 }  + {\bf{h}}_1^T {\bf{B}} {\bf{z}}_1  + z_D,
\end{equation}
where $x_{S_1 }$ and $x_{S_2 }$ are the transmitting signals from the two source nodes, ${\bf{h}}_{01}$ and ${\bf{h}}_{02}$ denote the broadcast channels between the two source nodes and relay nodes respectively, and ${\bf{h}}_{1}$ represents the multiple-access channel between the relay nodes and the destination node. The other parameters are defined in sections II and III. Again, assume each relay node has an individual transmitting power constraint (2). As in the point-to-point case, it is easy to verify that the equivalent constraint of the power constraint at each relay node $k$ in this network is given as follows.
\begin{equation}
\beta _k^2  \le \frac{{P_k^{Up} }}{{1 + h_{S_1 ,k}^2 P_{S_1 }  + h_{S_2 ,k}^2 P_{S_2 } }}\buildrel \Delta \over = \left( {\beta _k^{Up} } \right)^2, k=1,2 \cdots n.
\end{equation}
To distinguish a specific AF scheme $\bf{B}$, we use $\{\bf{B}\}$ to indicate all feasible relay amplification gains are allowed.

\begin{figure}
  \centering
  \includegraphics[width=2in,height=1.5in]{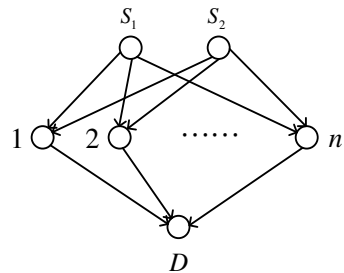}
  \caption{Gaussian two-hop multiple access channel}
\end{figure}

\newcounter{TempEqCnt50}
\setcounter{TempEqCnt50}{\value{equation}}
\setcounter{equation}{50}
\begin{figure*}[ht]
\begin{align}
{\cal{ R}}(\bf{B})  = & \left\{ {\left( {R_1 ,R_2 } \right):R_1  \le {\cal{C}}\left( {\frac{{{\bf{h}}_1^T {\bf{Bh}}_{01} {\bf{h}}_{01}^T {\bf{Bh}}_1 P_{S_1 } }}{{{\bf{h}}_1^T {\bf{BBh}}_1  + 1}}} \right)} \right., \nonumber\\
&\left. {R_2  \le {\cal{C}}\left( {\frac{{{\bf{h}}_1^T {\bf{Bh}}_{02} {\bf{h}}_{02}^T {\bf{Bh}}_1 P_{S_2 } }}{{{\bf{h}}_1^T {\bf{BBh}}_1  + 1}}} \right),R_1  + R_2  \le {\cal{C}}\left( {\frac{{{\bf{h}}_1^T {\bf{B}}{\bf{A}}{\bf{Bh}}_1 }}{{{\bf{h}}_1^T {\bf{BBh}}_1  + 1}}} \right)} \right\}\\
{\cal{ R}}^{out}_1(\bf{B})  = & \left\{ {\left( {R_1 ,R_2 } \right):R_1  \le {\cal{C}}\left( {\frac{{{\bf{h}}_1^T {\bf{Bh}}_{01} {\bf{h}}_{01}^T {\bf{Bh}}_1 P_{S_1 } }}{{{\bf{h}}_1^T {\bf{BBh}}_1  }}} \right)} \right., \nonumber\\
&\left. {R_2  \le {\cal{C}}\left( {\frac{{{\bf{h}}_1^T {\bf{Bh}}_{02} {\bf{h}}_{02}^T {\bf{Bh}}_1 P_{S_2 } }}{{{\bf{h}}_1^T {\bf{BBh}}_1  }}} \right),R_1  + R_2  \le {\cal{C}}\left( {\frac{{{\bf{h}}_1^T {\bf{B}}{\bf{A}}{\bf{Bh}}_1 }}{{{\bf{h}}_1^T {\bf{BBh}}_1  }}} \right)} \right\}\\
{\cal{R}}^{out}_2 \left( {\bf{B}} \right) = &\left\{ {\left( {R_1 ,R_2 } \right):R_1  \le {\cal{C}}\left( {{\bf{h}}_1^T {\bf{Bh}}_{01} {\bf{h}}_{01}^T {\bf{Bh}}_1 P_{S_1 } } \right)} \right., \nonumber\\
&\left. {R_2  \le {\cal{C}}\left( {{\bf{h}}_1^T {\bf{Bh}}_{02} {\bf{h}}_{02}^T {\bf{Bh}}_1 P_{S_2 } } \right),R_1  + R_2  \le {\cal{C}}\left( {{\bf{h}}_1^T {\bf{B}}{\bf{A}}{\bf{Bh}}_1 } \right)} \right\}
\end{align}
\hrulefill
\vspace*{4pt}
\end{figure*}
\setcounter{equation}{\value{TempEqCnt50}}

\newcounter{TempEqCnt54}
\setcounter{TempEqCnt54}{\value{equation}}
\setcounter{equation}{54}
\begin{figure*}[hb]
\hrulefill
\begin{align}
{\rm{col}} \left( {\rm{conv}}\left({\cal{R}}^{out}_2  \left( \{{\bf{B}}\} \right)\right)\right)  = &\left\{ {\left( {R_1 ,R_2 } \right):R_1  \le {\cal{C}}\left( {{\bf{h}}_1^T {\bf{B}}^{*}{\bf {h}}_{01} {\bf{h}}_{01}^T {\bf{B}}^{*}{\bf {h}}_1 P_{S_1 } } \right)} \right., \nonumber\\
&\left. {R_2  \le {\cal{C}}\left( {{\bf{h}}_1^T {\bf{B}}^{*}{\bf {h}}_{02} {\bf{h}}_{02}^T {\bf{B}}^{*}{\bf {h}}_1 P_{S_2 } } \right),R_1  + R_2  \le {\cal{C}}\left( {{\bf{h}}_1^T {\bf{B}}^{*}{\bf{A}}{{\bf{B}}^{*}{\bf {h}}}_1 } \right)} \right\}
\end{align}
\vspace*{4pt}
\end{figure*}
\setcounter{equation}{\value{TempEqCnt54}}

In the similar way, the received signal at the destination can be represented by an output of an equivalent scaler MAC shown as follows.
\begin{equation}
y_D = x_{S_1 ,eq}  + x_{S_2 ,eq}  + z_{eq},
\end{equation}
where $x_{S_1 ,eq}={\bf{h}}_1^T {\bf{B}} {\bf{h}}_{01} x_{S_1 }$ and $x_{S_2 ,eq}={\bf{h}}_1^T {\bf{B}} {\bf{h}}_{02} x_{S_2 }$ are two equivalent transmitting signals of the two sources respectively, and $z_{eq}={\bf{h}}_1^T {\bf{B}} {\bf{z}}_1  + z_D$ is the equivalent Gaussian noise. With a specific $\bf{B}$, the two-hop MAC can be considered as a conventional Gaussian MAC. To distinguish them, we denote the former one as MAC($\bf{B}$). As the well-known result of Gaussian MAC, the source nodes adopt the Gaussian codebooks. The r.v.'s used to generate the codebooks are $x_{S_1}\sim{\cal{N}}[0,P_{S_1}]$ and $x_{S_2}\sim{\cal{N}}[0,P_{S_2}]$. The codebooks consist of  $\left\lceil {2^{nR_1} } \right\rceil$ and  $\left\lceil {2^{nR_2} } \right\rceil$ codewords of length $n$ respectively, and the decoding error probability tends to zero as $n \to \infty$.

Corresponding to different AF schemes employed by the relay nodes, MAC$(\bf{B})$ may have different achievable rate sets denoted by ${\cal{R}}({\bf{B}})$ as shown in (51), where ${\bf{A}}={\bf{H}}_0 diag \{P_{S_1},P_{S_2}\}{\bf{H}}_0^T$, and ${\bf{H}}_0=[{\bf{h}}_{01},{\bf{h}}_{02}]$. Denote by ${\cal{R}}\left( \left\{{\bf{B}}\right\} \right)$ the union of all achievable rate sets. However, note that the union region may not be convex. Therefore, to obtain the entire achievable rate region, we consider the time-sharing technique according to different AF schemes. Consequently, the following corollary is established.

\newtheorem{col}{Corollary}
\begin{col}
The optimal achievable rate region of the two-hop MAC via AF relay scheme is the closure of the convex hull of all the possible rate sets (51),
\[
{\rm{col}} \left( {\rm{conv}}\left( {\cal{R}}\left( \left\{{\bf{B}}\right\} \right)\right)\right).
\]
\end{col}

The convex hull is taken by the time-sharing technique among all the rate sets. However, it is even more complicated to obtain the optimal achievable rate region of the two-hop MAC explicitly than to obtain the optimal rate as in the point-to-point case. So we focus on giving an asymptotical characterization of the rete region in the sequel.

\subsection{Outer Bound to Achievable Rate Region}
To investigate the structure of the optimal rate region, we first derive an outer bound for it in the similar manner as we obtain the
upper bound of the ANC rate for a layered relay network. If omitting the received noise at the destination node or the propagated
noises from the relay nodes, two larger rate sets ${\cal{R}}^{out}_1  \left( {\bf{B}} \right)$ and
${\cal{R}}^{out}_2  \left( {\bf{B}} \right)$ can also be obtained with AF scheme $\bf{B}$ respectively as shown in
(52) and (53), which yield an outer bound for the optimal rate region given in the following corollary.
\newtheorem{col2}{Corollary}
\begin{col}
An outer bound of the achievable rate region of the two-hop MAC via AF is the intersection of two outer bounds shown as follows.
\begin{center}
${\rm{col}} \left( {\rm{conv}}\left({\cal{R}}^{out}_1  \left( \{{\bf{B}}\} \right)\right)\right) \bigcap {\rm{col}} \left( {\rm{conv}}\left({\cal{R}}^{out}_2  \left( \{{\bf{B}}\} \right)\right)\right)$.
\end{center}
\end{col}

In the following, we wish to address the issue on deriving the two outer bounds, i.e., ${\rm{col}} \left( {\rm{conv}}\left({\cal{R}}^{out}_1  \left( \{{\bf{B}}\} \right)\right)\right)$
and ${\rm{col}} \left( {\rm{conv}}\left({\cal{R}}^{out}_2  \left( \{{\bf{B}}\} \right)\right)\right)$, in the closed forms. The first observation is that the second outer bound can be easily expressed in a closed form. It is easy to verify that the three Gaussian capacities in the rate set ${\cal{R}}^{out}_2(\bf{B})$ can be maximized simultaneously by a specific choice
\setcounter{equation}{53}
\begin{equation}
{\bf{B}}^{*}  = diag\left\{ {\beta _1^{Up}  \cdots \beta _k^{Up}  \cdots \beta _n^{Up} } \right\}.
\end{equation}
The corresponding rate set is obtained, which covers any other rate sets in ${\cal{R}}^{out}_2  \left( \{{\bf{B}}\} \right)$. Moreover, note that since ${\cal{R}}^{out}_2({\bf{B}}^{*})$ itself is a convex set, the second outer bound can be given in (55). However, to obtain the first outer bound is not such straightforward. Several useful lemmas are given as follows for characterizing it.
\newtheorem{lemma2}{Lemma}
\begin{lemma}
To have the first outer bound of the achievable rate region of a two-hop MAC, it is sufficient to take ${\bf{x}}={\bf{B}\bf{h}}_1$, which is called the amplification vector hereafter, in the linear subspace $span\left\{ {{\bf{h}}_{0i} ,i = 1,2} \right\}$.
\end{lemma}
\begin{IEEEproof}
The proof is given in appendix D.
\end{IEEEproof}

As shown in (52), the three Gaussian capacities are independent of the norm of $\bf{ x}$. By absorbing the norm of
${\bf{h}}_{01}$ and ${\bf{h}}_{02}$ into $P_{S_1 }$ and $P_{S_2 }$ respectively, it is assumed without loss of generality, that
${\bf{h}}_{0i}$, $i=1,2$ and $\bf{ x}$ are normalized vectors. Specifically, a two-hop MAC with two relay nodes is analyzed first.
By Lemma $2$, it is sufficient for us to work on a two-dimensional plane. We rewrite ${\bf{h}}_{0i}$, $i=1,2$ and $\bf{ x}$ as
\setcounter{equation}{55}
\begin{align}
{\bf{h}}_{01}  &= \left[ {\cos \alpha ,\sin \alpha } \right]^T,\\
{\bf{h}}_{02}  &= \left[ {\cos \beta ,\sin \beta } \right]^T,\\
{\bf{x}} & = \left[ {\cos \theta ,\sin \theta} \right]^T.
\end{align}

The key insight of the above expressions is that we can find a set of orthogonal bases of space
$span \left\{ {\bf{h}}_{0i}, i=1,2\right\}$, e.g.,  ${\bf{u}}_1= \left[1,0\right]^{T}$ and ${\bf{u}}_2= \left[0,1\right]^{T}$,
then project all the three vectors onto them. Therefore, from the geometric aspect, the elements in each vector can be interpreted as the coodinates under such base.
Assume without loss of generality that $\alpha \le \beta$ in the sequal.
By substituting (56)-(58) into (52), the rate set ${\cal{R}}^{out}_1(\bf{B})$ can be recast as
\begin{align}
\left\{ {\left( {R_1 ,R_2 } \right)} \right.&:R_1  \le {\cal{C}}\left( { \phi _1 \left( \theta  \right)} \right),\nonumber\\
&\left. {R_2  \le {\cal{C}}\left( { \phi _2 \left( \theta  \right)} \right),R_1  + R_2  \le {\cal{C}}( \phi \left( \theta  \right))} \right\},
\end{align}
where $\theta \in \left[0, 2\pi \right]$,
\begin{align}
 &\phi _1 \left( \theta \right) = P_{S_1 } \cos ^2 \left( {  \theta - \alpha} \right), \nonumber\\
 &\phi _2 \left( \theta \right) = P_{S_2 } \cos ^2 \left( {\theta - \beta} \right), \nonumber\\
 &\phi \left( \theta \right) = P_{S_1 } \cos ^2 \left( {\theta - \alpha} \right) + P_{S_2 } \cos ^2 \left( {\theta - \beta} \right). \nonumber
\end{align}

To emphasize this, later we use ${\cal{R}}^{ out }_1 \left( \theta  \right)$ to denote the rate set in the first outer bound and ${\cal{R}}^{out}_1 \left( \{ \theta \} \right)$ to denote the union of them. With the assistant of the above expression, we investigate the relationship between different rate sets in the first outer bound and find an interesting result with which the scope of $\theta$ to be considered is narrowed to $\left[ \alpha , \beta \right]$.

\newtheorem{lemma3}{Lemma}
\begin{lemma}
To have the first outer bound of the achievable rate region of a two-hop MAC with two relay nodes, it is sufficient to take $\theta \in [\alpha , \beta]$.
\end{lemma}
\begin{IEEEproof}
The proof is given in appendix E.
\end{IEEEproof}

Therefore, we only consider ${\cal{R}}^{out}_1 \left( \{ \theta \} \right)$ for $\theta \in [\alpha , \beta]$ in the sequel. There still another problem remains to be solved. Is the time-sharing operation necessary? In other words, is the union region is a convex one? From (59), it shows that if both $\phi_i \left( \theta \right)$, $i=1,2$ are concave, then $\phi \left( \theta \right)$ is concave, and  since $\cal{C} \left( \cdot \right)$ is an increasing concave function, the three capacities in (59) are all concave functions of $\theta$. Consequently, the union of all rate sets ${\cal{R}}^{out}_1 \left( \{ \theta \} \right)$ is a convex set. In this case, ${\rm{conv}}\left( {\cal{R}}^{out}_1 \left( \{ \theta \} \right) \right) = {\cal{R}}^{out}_1 \left( \{ \theta \} \right)$. In the following lemma, a sufficient condition for this case is proposed.

\newtheorem{lemma4}{Lemma}
\begin{lemma}
The first outer bound of the achievable rate region of a two-hop MAC with two relay nodes is convex when $\left| {\alpha  - \beta } \right| \le \frac{\pi }{4}$.
\end{lemma}
\begin{IEEEproof}
By Lemma 3, we only consider $\theta \in [\alpha, \beta]$. Hence if $\left| {\alpha  - \beta } \right| \le \frac{\pi }{4}$, then $\left| {\theta- \alpha } \right| \le \frac{\pi }{4}$ and $\left| {\theta  - \beta } \right| \le \frac{\pi }{4}$. From (59) we see that
\begin{equation}
\frac{{d^2 }}{{d\theta^2 }}\phi _1 \left( \theta \right) =  - 2P_{S_1 } \cos 2\left( {\theta - \alpha } \right) \le 0
\end{equation}
when $\left| {\theta- \alpha } \right| \le \frac{\pi }{4}$, which implies $\phi_1 \left( \theta \right)$ is concave, and
\begin{equation}
\frac{{d^2 }}{{d\theta^2 }}\phi _2 \left( \theta \right) =  - 2P_{S_2 } \cos 2\left( {\theta - \beta } \right) \le 0
\end{equation}
when $\left| {\theta- \beta } \right| \le \frac{\pi }{4}$, which implies $\phi_2 \left( \theta \right)$ is concave.

Consequently,  $\phi \left( \theta \right)$ is concave, and thus ${\cal{R}}^{out}_1 \left( \{ \theta \} \right)$ is convex.
\end{IEEEproof}

Indeed, each result obtained in the above lemmas can be extended to a two-hop MAC with $n$ relay nodes directly.
The key step is the observed in proof of the following theorem.
\newtheorem{thm5}{Theorem}
\begin{thm}
To have the first outer bound of the achievable rate region of a two-hop MAC with arbitrary $n$ relay nodes, it is sufficient to take $\theta \in [\alpha , \beta]$. If $|\arccos ({\bf{h}}_{01}^T{\bf{h}}_{02})| \le \frac{\pi}{4}$, then ${\cal{R}}^{out}_1 \left( \{ \theta \} \right)$ is a convex set and hence
\begin{equation}
{\rm{conv}}\left( {\cal{R}}^{out}_1 \left( \{ \theta \} \right) \right) = {\cal{R}}^{out}_1 \left( \{ \theta \} \right), \theta \in [\alpha , \beta].
\end{equation}
\end{thm}
\begin{IEEEproof}
The proof is given in appendix F.
\end{IEEEproof}

However, the structure of the first outer bound is still unknown in general case. We progressively work towards this goal. A commonly
used method to characterize different rate-tuples on the boundary of a multiuser capacity region is via solving a sequence of weighted
sum rate maximization problems, each for a different nonnegative rate weight vector of users (see [14] and also [21]). For a two-hop
MAC, such technique can slao be employed to characterize ${\cal{R}}^{out}_1 \left( \{ \theta \} \right)$. That is to solve an
optimization problem ${\max }  \mu _1 R_1  + \mu _2 R_2 ,{\rm{   }}\mu _1 ,\mu _2  \ge 0$. It is clear that the
rate pair $\left( {R_1 ,R_2 } \right)$ on the boundary of the union region must be on the boundary of some rate region
${\cal{R}}^{out}_1(\theta)$, which should satisfy (52). By such observation, we first analyze the relationship between the boundary points of
${\cal{R}}^{out}_1(\theta)$ and the given weight vector $(\mu_1 , \mu_2)$ in the optimization problem. Without loss of generality, assume that $\mu_1$ and $\mu_2$ are normalized in
$\left[ 0,1\right]$, and that $\mu_1 + \mu_2=1$. Therefore, the maximization can be recast as
${\max } \mu R_1  + \bar{\mu} R_2$, $\mu=\mu _1 ,\bar{\mu}=\mu _2$. For any given $\theta \in \left[\alpha, \beta\right]$, it is easy to verify that


\newcounter{TempEqCnt62}
\setcounter{TempEqCnt62}{\value{equation}}
\setcounter{equation}{62}
\begin{figure*}
\begin{equation}
\left\{
\begin{aligned}
\theta^{*}(\frac{1}{2})  &= \frac{1}{2}\arctan \left( {\frac{{P_{S_1 } \sin 2\alpha  + P_{S_2 } \sin 2\beta }}{{P_{S_1 } \cos 2\alpha  + P_{S_2 } \cos 2\beta }}} \right),&\frac{{P_{S_1 } \sin 2\alpha  + P_{S_2 } \sin 2\beta }}{{P_{S_1 } \cos 2\alpha  + P_{S_2 } \cos 2\beta }} \ge 0 \\
\theta^{*}(\frac{1}{2})  &= \frac{1}{2}\left[ {\pi  + \arctan \left( {\frac{{P_{S_1 } \sin 2\alpha  + P_{S_2 } \sin 2\beta }}{{P_{S_1 } \cos 2\alpha  + P_{S_2 } \cos 2\beta }}} \right)} \right],&\frac{{P_{S_1 } \sin 2\alpha  + P_{S_2 } \sin 2\beta }}{{P_{S_1 } \cos 2\alpha  + P_{S_2 } \cos 2\beta }} \le 0
\end{aligned}
\right.
\end{equation}
\hrulefill
\vspace*{4pt}
\end{figure*}
\setcounter{equation}{\value{TempEqCnt62}}

\newcounter{TempEqCnt67}
\setcounter{TempEqCnt67}{\value{equation}}
\setcounter{equation}{67}
\begin{figure*}[ht]
\begin{align}
 \mu &\left[ {P_{S_1 } P_{S_2 } \sin 2\left( {\beta  - \theta } \right)\cos ^2 \left( {\theta  - \alpha } \right) + P_{S_1 } \sin 2\left( {\theta  - \alpha } \right)\left( {1 + P_{S_2 } \cos ^2 \left( {\theta  - \beta } \right)} \right)} \right] \nonumber\\
  &= \bar{\mu} \left[ {P_{S_1 } P_{S_2 } \sin 2\left( {\beta  - \theta } \right)\cos ^2 \left( {\theta  - \alpha } \right) + P_{S_2 } \sin 2\left( {\beta  - \theta } \right)\left( {1 + P_{S_2 } \cos ^2 \left( {\theta  - \beta } \right)} \right)} \right]
\end{align}
\begin{align}
 \mu &\left[ {P_{S_1 } P_{S_2 } \sin 2\left( {\theta  - \alpha } \right)\cos ^2 \left( {\theta  - \beta } \right) + P_{S_1 } \sin 2\left( {\theta  - \alpha } \right)\left( {1 + P_{S_1 } \cos ^2 \left( {\theta  - \alpha } \right)} \right)} \right] \nonumber\\
  &= \bar{\mu} \left[ {P_{S_1 } P_{S_2 } \sin 2\left( {\theta  - \alpha } \right)\cos ^2 \left( {\theta  - \beta } \right) + P_{S_2 } \sin 2\left( {\beta  - \theta } \right)\left( {1 + P_{S_1 } \cos ^2 \left( {\theta  - \alpha } \right)} \right)} \right]
\end{align}
\hrulefill
\vspace*{4pt}
\end{figure*}
\setcounter{equation}{\value{TempEqCnt67}}

\begin{itemize}
\item \textbf{case 1}: when $\mu = 1/2 $, the rate pairs maximize the sum rate in rate set
      ${\cal{ R}}^{out}_1 \left( \theta  \right) $ are the solutions of the optimization problem, which satisfy
    \begin{center}
    $R_1(\theta) +R_2(\theta) = {\cal{C}}\left( {\phi \left( {\theta } \right)} \right)$.
    \end{center}
\item \textbf{case 2}: when $\mu = 1 $, the rate pairs maximize the individual rate $R_1$ in rate
      set ${\cal{ R}}^{out}_1 \left( {\theta}  \right) $ are the solutions of the optimization problem, which satisfy
    \begin{center}
     $R_1(\theta)  = {\cal{C}}\left( {\phi _1 \left( {\theta } \right)} \right)$,
     $0 \le R_2 (\theta) \le {\cal{C}}\left( {\phi \left( {\theta } \right)} \right) - {\cal{C}}\left( {\phi _1 \left( {\theta } \right)} \right)$.
    \end{center}
\item \textbf{case 3}: when $\mu= 0$, the rate pairs maximize the individual rate $R_2$ in rate set
      ${\cal{ R}}^{out}_1 \left( \theta   \right) $ are the solutions of the optimization problem, which satisfy
    \begin{center}
    $0 \le R_1(\theta)  \le {\cal{C}}\left( {\phi \left( {\theta } \right)} \right)-{\cal{C}}\left( {\phi_2 \left(  {\theta  } \right)} \right)$,
    $R_2 (\theta) = {\cal{C}}\left( {\phi_2 \left( {\theta  } \right)} \right)$.
    \end{center}
\item \textbf{case 4}: when $\mu \in \left( 0, \frac{1}{2}\right)$, the rate pair at the "upper-diagonal" corner point of the
      pentagon region (see Fig. 13) in rate set ${\cal{ R}}^{out}_1 \left( \theta  \right) $ is the solution of the optimization
      problem, i.e.,
    \begin{center}
    $R_1 (\theta) = {\cal{C}}\left( {\phi \left( \theta  \right)} \right) - {\cal{C}}\left( {\phi _2 \left(  \theta \right)} \right)$,
    $R_2 (\theta) = {\cal{C}}\left( {\phi _2 \left(  \theta  \right)} \right)$.
    \end{center}
\item \textbf{case 5}: when $\mu \in \left(\frac{1}{2}, 1\right)$, the rate pair at the "lower-diagonal" corner point of the
      pentagon region (see Fig. 13) in rate set ${\cal{ R}}^{out}_1 \left( \theta  \right)$  is the solution of the optimization
      problem, i.e.,
    \begin{center}
    $R_1(\theta)  = {\cal{C}}\left( {\phi_1 \left( \theta  \right)} \right) $,
    $R_2 (\theta) = {\cal{C}}\left( {\phi \left( \theta  \right)} \right)- {\cal{C}}\left( {\phi _1 \left( \theta  \right)} \right)$.
    \end{center}
\end{itemize}

\begin{figure}
  \centering
  \includegraphics[width=2in]{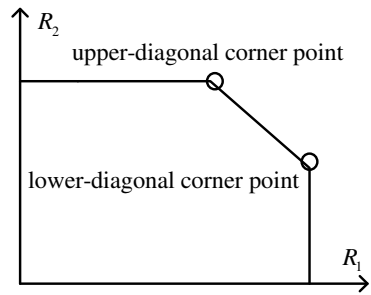}
  \caption{Rate set in first outer bound}
\end{figure}

With the analysis made above, we conclude that for any given $\mu$, the boundary points of ${\cal{ R}}^{out}_1 \left(\left\{ \theta  \right\}\right)$
can be derived by miximizing $ \mu R_1(\theta)  + \bar{\mu } R_2(\theta)$ with respect to $\theta$ and the optimal solution is denoted by $\theta^{*}(\mu)$.
We explicitly obtain the rate pairs on the boundary which maximize
the sum rate and the individual rates of ${\cal{ R}}^{out}_1 \left(\left\{ \theta  \right\}\right)$ in the sequel. Then two equations
will be presented, whose solutions maximize $\mu R_1 (\theta) + \bar{\mu} R_2(\theta)$ for $\mu \in \left( 0, \frac{1}{2}\right)$ and
$\mu \in \left(\frac{1}{2}, 1\right)$ respectively.

Let us start with the maximum sum rate of ${\cal{R}}^{out}_1 \left( \{ \theta \} \right)$. The following lemma is established.
\newtheorem{lemma5}{Lemma}
\begin{lemma}
The maximum value of $R_1 + R_2$ of ${\cal{ R}}^{out}_1 \left(\left\{ \theta  \right\}\right)$ is given by ${\cal{C}}\left( {\phi {\rm{ }}\left( {\theta^{*}(\frac{1}{2}) } \right)} \right)$, where $\theta^{*}(\frac{1}{2})$ is shown in (63) on the top of the next page.
\end{lemma}
\begin{IEEEproof}
For any $\theta$, the maximum sum rate $R_1+R_2$ is upper bounded by ${\cal{C}}\left(\phi\left( {\theta  } \right)\right)$. It corresponds to the maximization problem when $\mu = 1/2 $ as shown in case 1.
By setting the derivative $\phi '\left( {\theta  } \right)$ to zero, we have
\setcounter{equation}{63}
\begin{equation}
\phi '\left( {\theta  } \right) =  - P_{S_1 } \sin 2\left( {\theta   - \alpha } \right) - P_{S_2 } \sin 2\left( {\theta  - \beta } \right) = 0.
\end{equation}
Then, it follows that the solution of the above equation satisfies
\begin{equation}
\tan 2\theta^{*}(\frac{1}{2})  = \frac{{P_{S_1 } \sin 2\alpha  + P_{S_2 } \sin 2\beta }}{{P_{S_1 } \cos 2\alpha  + P_{S_2 } \cos 2\beta }}.
\end{equation}
Since $\theta ^{*}(\frac{1}{2})  \in \left[ {\alpha ,\beta } \right]$, $\theta^{*}(\frac{1}{2})$ is obtained as shown in (63) on the top of the page.
The AF scheme corresponging to $\theta^{*}(\frac{1}{2})$ is denoted by ${\bf{B}}^{(10)}$. By the notations used in the proof of Theorem 5,
\[{\bf{B}}^{(10)}=c^{(10)} diag \left[{\bf{H}}^{-1}_1\left( \cos \theta^{*}(\frac{1}{2}) {\bf{u}}_1+\sin \theta^{*}(\frac{1}{2}) {\bf{u}}_2\right)\right],\]
where the constant $c^{(10)}$ is chosen such that the constraint (49) is satisfied and at least one of the amplification gain achieves the upper bound.
\end{IEEEproof}

Then we consider the maximum individual rates of ${\cal{R}}^{out}_1 \left( \{ \theta \} \right)$. The following lemma is established.

\newtheorem{lemma6}{Lemma}
\begin{lemma}
The maximum value of individual rates $R_i$, $i=1,2$ of ${\cal{ R}}^{out}_1 \left(\left\{ \theta  \right\}\right)$ are given by ${\cal{C}}\left( {\phi_1 \left( {\alpha } \right)} \right)$ and ${\cal{C}}\left( {\phi_2 \left( {\beta } \right)} \right)$.
\end{lemma}
\begin{IEEEproof}
To derive the maximum individual rates is equivalent to solving the maximization problems ${\max}{\cal{C}}(\phi_1 \left( {\theta  } \right))$
and  ${\max}{\cal{C}}(\phi_2 \left( {\theta  } \right))$ according to $\mu=1$ and $\mu=0$ respectively as shown in case 2 and 3.
By setting the derivative $\phi_1 '\left( {\theta  } \right)$ to zero, we have
\begin{equation}
\phi_1 '\left( {\theta } \right) =  - P_{S_1 } \sin 2\left( {\theta   - \alpha } \right)=0.
\end{equation}
The solution of the above equation is $\theta^{*}(1)=\alpha$.

By setting the derivative $\phi_2 '\left( {\theta  } \right)$ to zero, we have
\begin{equation}
 \phi_2 '\left( {\theta } \right) =- P_{S_2 } \sin 2\left( {\theta  - \beta } \right) = 0.
\end{equation}
The solution of the above equation is $\theta^{*}(0)=\beta$.

By the notations used in the proof of Theorem 5, the AF schemes corresponging to $\theta^{*}(1)$ and $\theta^{*}(0)$ are denoted by
${\bf{B}}^{(11)}=c^{(11)} diag \left[ {\bf{H}}^{-1}_1\left( \cos \theta^{*}(1) {\bf{u}}_1+\sin \theta^{*}(1) {\bf{u}}_2\right)\right]$, and
${\bf{B}}^{(12)}=c^{(12)} diag \left[{\bf{H}}^{-1}_1\left( \cos \theta^{*}(0) {\bf{u}}_1+\sin \theta^{*}(0) {\bf{u}}_2\right)\right]$, where the
constant $c^{(1i)}, i=1,2$ is chosen such that the constraint (49) is satisfied and at least one of the amplification gain in ${\bf{B}}^{(1i)}$
achieves the upper bound.
\end{IEEEproof}

\newtheorem{lemma7}{Lemma}
\begin{lemma}
The weighted sum rate $\mu R_1(\theta)  + \bar{ \mu} R_2(\theta)$ is maximized by $\theta^{*}(\mu)$ satisfying (68) given on the top of the page for $\mu \in \left(0,\frac{1}{2}\right)$,
and is maximized by $\theta^{*}(\mu)$ satisfying (69) given on the top of the page, for $\mu \in \left(\frac{1}{2}, 1 \right)$.
\end{lemma}
\begin{IEEEproof}
Since for $\mu \in \left(0,\frac{1}{2}\right)$, the rate pair $(R_1,R_2)$ on the boundary can be determined by solving the maximization problem $\mu \left({\cal{C}}\left( {\phi \left( \theta  \right)} \right) - {\cal{C}}\left( {\phi _2 \left( \theta  \right)} \right)\right)  + \bar{\mu}{\cal{C}}\left( {\phi _2 \left( \theta  \right)} \right)$. By setting the derivative of the object function to zero, we obtain (68).

By symmetry, the proof of the second part is exactly the same as the first part.
\end{IEEEproof}

\begin{figure}
  \centering
  \includegraphics[width=2in]{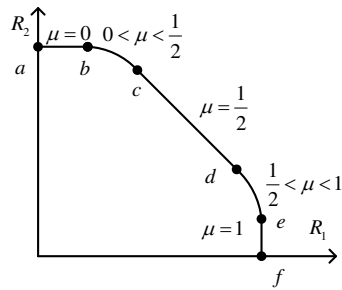}
  \caption{First outer bound}
\end{figure}

From (68), it is clear that each $\theta^*(\mu)$, $\mu \in (0,\frac{1}{2})$ is in $ \left[ {\theta ^*(\frac{1}{2}) ,\beta } \right]$
and conversely for each $ \theta \in \left[\theta^*(\frac{1}{2}),\beta \right]$ there exists a $\mu \in (0,\frac{1}{2})$ satisfying (68).
Consequently, the boundary points determined by the optimization problem with $\mu \in (0,\frac{1}{2})$ (see Fig. 14, $bc$) can also be
determined by the parametric function of the "upper-diagonal" corner point of the region corresponding to the rate set
${\cal{R}}^{out}_1(\theta)$ with $\theta  \in \left[ {\theta ^{*}(\frac{1}{2}) ,\beta } \right]$. Similarly as the previous case, it is
clear that each $\theta^*(\mu) $ for given $\mu \in (\frac{1}{2},1)$ is in $\left[ {\alpha, \theta ^{*}(\frac{1}{2})  } \right]$
and conversely for each $ \theta \in \left[\alpha,\theta ^{*}(\frac{1}{2}) \right]$ there exists a $\mu \in (\frac{1}{2},1)$ satisfying (69).
Therefore, we claim that the boundary points determined by the optimization problem with $\mu \in (\frac{1}{2},1)$ (see Fig. 14, $cd$) can be
determined by the parametric function of the "lower-diagonal" corner point of the region corresponding to the rate set
${\cal{R}}^{out}_1(\theta)$ with $\theta  \in \left[ {\alpha, \theta ^{*}(\frac{1}{2})  } \right]$.

Finally, the following lemma is proposed for the purpose of analyzing the convexity of ${\cal{R}}^{out}_1 \left( \{ \theta \} \right)$.

\newtheorem{lemma8}{Lemma}
\begin{lemma}
The parametric function of the "upper-diagonal" corner point is a concave function of $\theta$, for $\theta \in \left[\theta^{*}(\frac{1}{2}), \beta\right]$, and the parametric function of the "lower-diagonal" corner point is a concave function of $\theta$, for $\theta \in \left[\alpha , \theta^{*}(\frac{1}{2}) \right]$.
\end{lemma}
\begin{IEEEproof}
The proof is presented in Appendix G.
\end{IEEEproof}

By the lemmas at hand, the entire region of the first outer bound can be fully characterized and thus the outer bound.

\subsection{Dynamic AF-Based ANC Scheme}
We focus on designing a practical AF-based ANC scheme in the left of the paper. Consider the three Gaussian capacity formulas in the achievable rate set ${\cal{R}}({\bf{B}})$, we find that each of them corresponds to a point-to-point two-hop relay network discussed in section V. According to different scenarios, differently many AF schemes are stored in the relay nodes constituting a dynamic ANC scheme. By employing the time-sharing technique, the achievable rate region of the dynamic ANC scheme is the closure of the convex hull of the rate sets it contains. It is obvious that one can obtain a larger rate region by storing more AF schemes in the relay nodes, but it will increase the complexity of the relaying scheme. So the practical ANC design involves a trade-off between the performance and complexity.

In the following, we consider four typical AF schemes. The first three ones are ${\bf{B}}^{(10)}$,
${\bf{B}}^{(11)}$ and ${\bf{B}}^{(12)}$. The forth AF scheme considered, denoted by ${\bf{B}}^{(2)}$, consists of the amplification gains as the upper bounds of the constraints (49). Similarly,
the upper layer noises under different AF schemes are denoted by $w^{(1i)}_D$, $i=0,1,2$ and $w^{(2)}_D$. We redefine the domination of
the noises received at the destination.

\newtheorem{definition6}{Definition}
\begin{definition}
We call the upper layer noise dominates the total noises received at the destination in a two-hop relay network with AF relays when \[
E\left[(w^{(2)}_D)^2\right] \ge E\left[(w^{(1i)}_D)^2\right] \ge 1, i=0,1,2
.\] Reversely, we call the lower layer noise dominates the total noises when
\[1 \ge E\left[(w^{(2)}_D)^2\right] \ge E\left[(w^{(1i)}_D)^2\right], i=0,1,2.\]
\end{definition}

Under these two special scenarios, the performence of two dynamic ANC schemes are analyzed.
\newtheorem{thm6}{Theorem}
\begin{thm}
For a two-user two-hop MAC with AF relays under individual power constraints, the gap between the achievable rate region and the optimal AF rate region is at most $1/2$ bit with respect to the maximum individual rates and the sum rate when either the upper layer noise or the lower layer noise dominates the total noises.
\end{thm}
\begin{IEEEproof}
For the first scenario, we compute the achieveble rate region with a dynamic ANC scheme consisting of three schemes, i.e., ${\bf{B}}^{(10)}$, ${\bf{B}}^{(11)}$ and ${\bf{B}}^{(12)}$.
The corresponding achievable rate region is the closure of the convex hull of ${\cal{R}}({\bf{B}}^{(1i)})$, $i=0,1,2$. For the gap between the outer bound and the maximum achievable
individual rates, the proof is exactly the same as the point-to-point case given in Theorem 3. For the sake of brevity, we omit the detailed proof here.

Denote the maximum sum rate of the optimal AF region by $(R_1+R_2)_{max}$. We need to show that the maximum sum rate of the achievable rate region is at most $1/2$ bit from $(R_1+R_2)_{max}$. From (51), the sum rate corresponding to scheme ${\bf{B}}^{(12)}$ is given as follows.
\begin{align}
R_1  + R_2 & = \frac{1}{2} \log \left( 1+{\frac{{{\bf{h}}_1^T {\bf{B}}^{(12)}{\bf{A}}{\bf{B}}^{(12)}{\bf{h}}_1 }}{{{\bf{h}}_1^T {\bf{B}}^{(12)}{\bf{B}}^{(12)}{\bf{h}}_1  + 1}}} \right)\nonumber\\
& = \frac{1}{2}\log \left( {1 + \frac{{{{{\bf{h}}_1^T {\bf{B}}^{\left( 12 \right)} {\bf{A}} {\bf{B}}^{\left( 12 \right)} {\bf{h}}_1 } \mathord{\left/
 {\vphantom {{{\bf{h}}_1^T {\bf{B}}^{\left( 12 \right)} {\bf{A}} {\bf{B}}^{\left( 12 \right)} {\bf{h}}_1 } {E\left[ {(w_D^{(12)} )^2 } \right]}}} \right.
 \kern-\nulldelimiterspace} {E\left[ {(w_D^{(12)} )^2 } \right]}}}}{{1 + {1 \mathord{\left/
 {\vphantom {1 {E\left[ {(w_D^{(12)} )^2 } \right]}}} \right.
 \kern-\nulldelimiterspace} {E\left[ {(w_D^{(12)} )^2 } \right]}}}}} \right) \nonumber
\end{align}
\setcounter{equation}{69}
\begin{align}
&\mathop  \ge \limits^{\left( a \right)} \frac{1}{2}\log \left( {1 + \frac{{{{{\bf{h}}_1^T {\bf{B}}^{\left(1 2 \right)} {\bf{A}} {\bf{B}}^{\left( 12 \right)} {\bf{h}}_1 } \mathord{\left/
 {\vphantom {{{\bf{h}}_1^T {\bf{B}}^{\left( 12 \right)} {\bf{A}}{\bf{B}}^{\left( 12 \right)} {\bf{h}}_1 } {E\left[ {(w_D^{(12)} )^2 } \right]}}} \right.
 \kern-\nulldelimiterspace} {E\left[ {(w_D^{(12)} )^2 } \right]}}}}{2}} \right) \nonumber\\
 &\ge \frac{1}{2}\log \left( {1 + \frac{{{\bf{h}}_1^T {\bf{B}}^{\left(1 2 \right)} {\bf{A}} {\bf{B}}^{\left( 12 \right)} {\bf{h}}_1 }}{{E\left[ {(w_D^{(12)} )^2 } \right]}}} \right) - \frac{1}{2} \nonumber\\
 &\mathop  \ge \limits^{\left( b \right)} (R_1+R_2)_{max} - \frac{1}{2},
\end{align}
where (a) follows from the assumption that $E\left[(w^{(12)}_D)^2\right] \ge 1$ and (b) follows from Lemma 5 that the maximum sum rate
of the optimal achievable rate region is no larger than the one of the first outer bound, i.e.,

\begin{align}
(R_1+&R_2)_{max} \nonumber\\
&\le {\cal{C}}(\phi(\theta^*(\frac{1}{2})))= \frac{1}{2}\log \left( {1 + \frac{{{\bf{h}}_1^T {\bf{B}}^{\left(1 2 \right)}
 {\bf{A}} {\bf{B}}^{\left( 12 \right)} {\bf{h}}_1 }}{{E\left[ {(w_D^{(12)} )^2 } \right]}}} \right).\nonumber
\end{align}

For the second scenario, the dynamic ANC scheme only consists of the forth scheme ${\bf{B}}^{(2)}$. The proof is almost the same as the previous case. Therefore, the proof is skipped here.
\end{IEEEproof}

To be more general, we can obtain a more complicated ANC scheme which is not limited to consisting of the above four AF schemes. It is clear, the more AF schemes stored at the relay nodes the lager the achievable rate region we may get. Hence, according to different application goals, we can have a tradeoff between the complexity of the coding scheme and the performance of it as the common argument.

\begin{figure}
  \centering
  \includegraphics[width=2in]{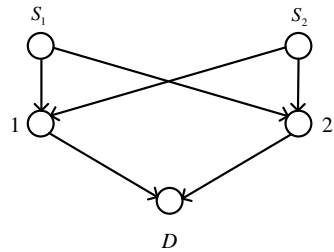}
  \caption{Two-hop MAC with two relay nodes}
\end{figure}
\subsection{Examples}
A two-hop MAC with two relay nodes is depicted in Fig. 14. Consider the general case and the two cases either the upper layer noise or the lower layer noise dominates the total noises received at the destination node. The corresponding network parameters are given in Figs. 15-17.

In the first case, we employ an ANC scheme with the four AF schemes ${\bf{B}}^{(1i)},i=0,1,2$ and ${\bf{B}}^{(2)}$ and the time-sharing technique. We explicitly plot the four rate regions in Fig. 15. The numerical result shows that the gap between the inner and outer bounds is approximate to $1/2$ bit.

In the second case, we employ an ANC scheme with the four AF schemes ${\bf{B}}^{(1i)},i=0,1,2$ and the time-sharing technique. Since $E\left[(w^{(2)}_D)^2\right]=E\left[(w^{(10)}_D)^2\right]=6.33$, and $E\left[(w^{(11)}_D)^2\right]=E\left[(w^{(12)}_D)^2\right]=3.29$, the upper layer noises dominate the total noises. Therefore the inner bound obtained by the ANC scheme yields a $1/2$ bit gap to the outer bound for the sum rate and two individual rates as shown in Fig. 16.

In the last case, only the forth AF scheme is employed. Since $E\left[(w^{(2)}_D)^2\right]=E\left[(w^{(10)}_D)^2\right]=0.38$, and $E\left[(w^{(11)}_D)^2\right]=E\left[(w^{(12)}_D)^2\right]=0.20$, the lower layer noise dominates the total noises. Therefore, the inner bound obtained by the ANC scheme yields a $1/2$ bit gap to the outer bound for the sum rate and two individual rates as shown in Fig. 17.

\begin{figure}
  \centering
  \includegraphics[width=3.5in]{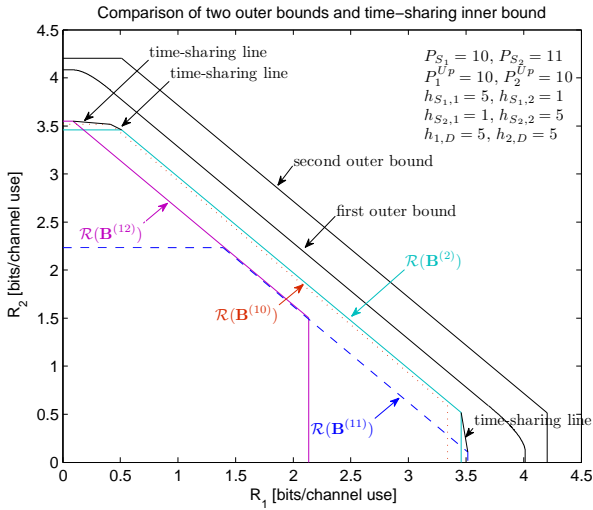}
  \caption{Time-sharing inner bound with AF schemes 1-4 and outer bound}
\end{figure}
\begin{figure}
  \centering
  \includegraphics[width=3.5in]{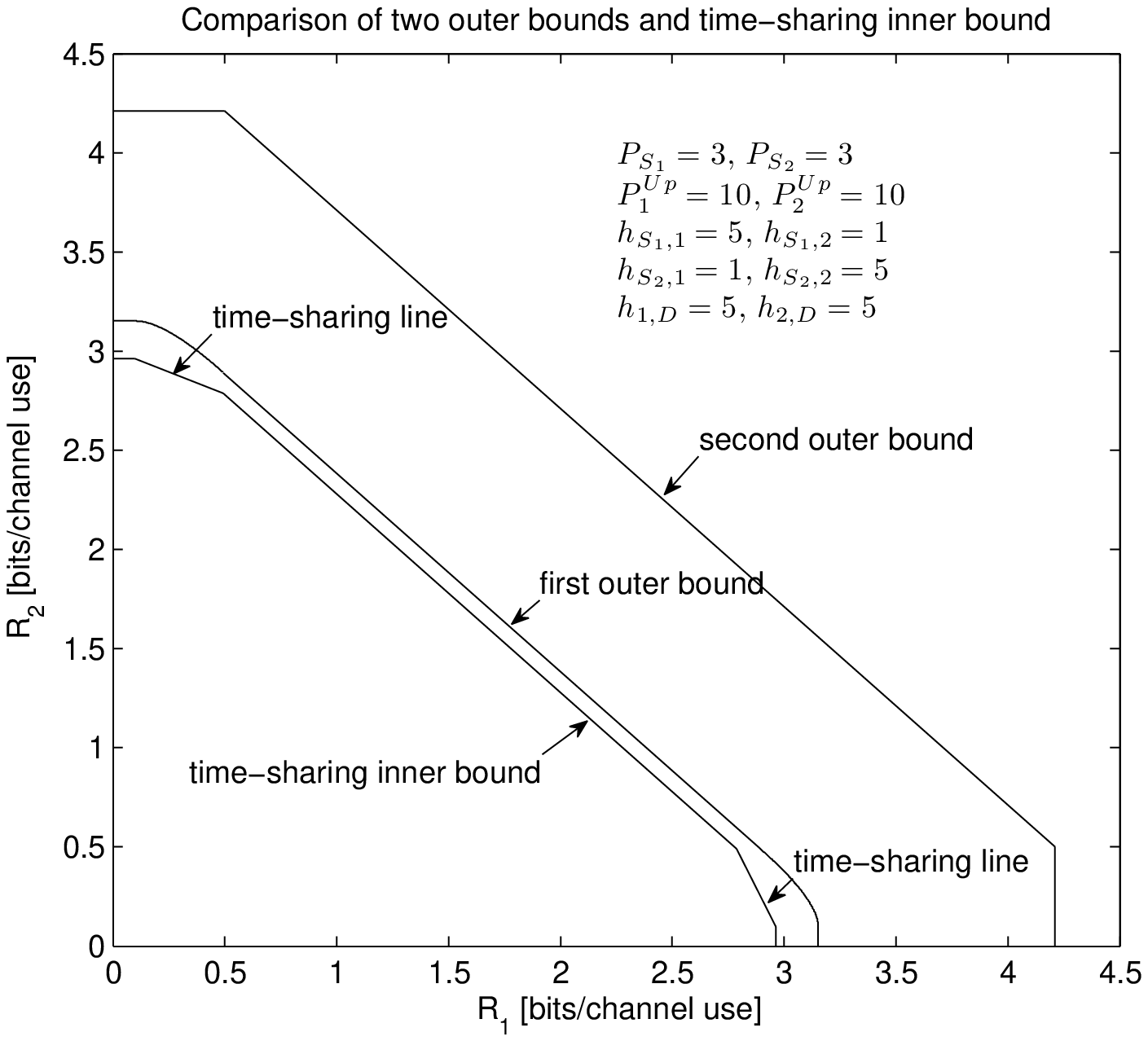}
  \caption{Time-sharing inner bound with AF schemes 1-3 and outer bound}
\end{figure}
\begin{figure}
  \centering
  \includegraphics[width=3.3in]{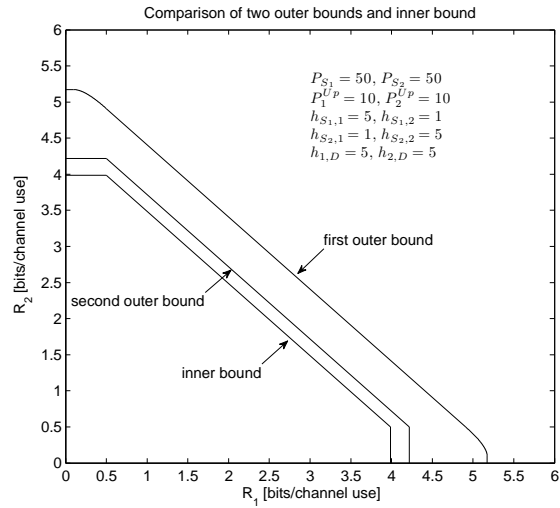}
  \caption{Inner bound with AF scheme 4 and outer bound}
\end{figure}

\section{Conclusion}
We cast the problem of computing the maximum achievable rate of AF-based ANC scheme as an optimization problem. For a multi-hop wireless relay network, assuming an individual power constraint at each relay node, we derive a tighter amplify-and-forward upper bound for the optimal transmission rate. The closed-form of such bound is derived for a layered network with a single source-destination pair. We have provided the analysis of the AF-based ANC scheme both for a point-to-point case and a two-hop MAC case with the assumption of instantaneous relay operation. However, in general wireless relay networks with AF relaying, the resulting input-output channel between the source and the destination is an intersymbol interference (ISI) channel with colored noise. We want to extend the results in this paper to that general case, and we conjecture the corresponding ANC scheme may be much more similar to the conventional network coding framework.


%

\appendices
\section{Proof of Lemma 1}
\begin{IEEEproof}
Let $w_k$ be the total noise received at node k.
\setcounter{equation}{70}
\begin{equation}
w_k  = \sum\limits_{\left\{ {i_k } \right\}} {f_{i_k ,k} z_{i_k } }  + z_k,
\end{equation}
where $\{i_k\}$ denotes the nodes appearing in the relaying path from $S$ to $k$. The signal received at node $k$ is expressed as
\begin{equation}
y_k=f_{S,k}x_S+w_k
\end{equation}
We prove the theorem by induction. Consider first the node k whose neighboring nodes set only contains the source node
S. From (1) and (3),
\begin{equation}
E\left[ {x_k^2 } \right] = \beta _k^2 E\left[ {y_k^2 } \right]\mathop  \le \limits^{\left( a \right)} \frac{{P_k^{Up}
}}{{\left( {1 + \delta _k } \right)P_{R,k} }}\left( {h_{S,k}^2 P_{\rm{S}}  + 1} \right)\mathop  = \limits^{\left( b
\right)} P_k^{Up},
\end{equation}
where (a) is obtained from the condition (11), and (b) is drawn from (9).
To prove that the theorem holds for any node, we assume that all the neighboring nodes of node $k$ satisfy the power constraints such that
\begin{equation}
E\left[ {x_j^2 } \right] = E\left[ {\beta _j^2 \left( {f_{S,j} x_S  + W_j } \right)^2 } \right] \le P_j^{Up} ,j \in {\cal{V}}\left( k
\right)
\end{equation}
Then we consider the transmitting power at node $k$ with the amplification gain chosen satisfying the condition (11),
\begin{align}
E\left[ {x_k^2 } \right]&\mathop = \limits^{\left( a \right)} E\left[ {\beta _k^2 \left( {f_{S,k} x_S  + \sum\limits_{j
\in {\cal{V}}\left( k \right)} {\beta _j h_{j,k} w_j }  + z_k } \right)^2 } \right]\nonumber\\
&\mathop  \le \limits^{\left( b \right)} c_{00}\left\{ {f_{S,k}^2 P_S  + E\left[ {\left( {\sum\limits_{j \in {\cal{V}}\left( k
\right)} {\beta _j h_{j,k} W_j } } \right)^2 } \right]+1} \right\}\nonumber\\
&\mathop  = \limits^{\left( c \right)}c_{00}\left\{ {c_{01}  + \sum\limits_{j \in {\cal{V}}\left( k \right)} {h_{j,k}^2 \beta
_j^2 \left[ {f_{S,j}^2 P_S  + E\left( {w_j^2 } \right)} \right]} } +1\right\} \nonumber\\
&\mathop  \le \limits^{\left( d \right)} c_{00}\left\{ {c_{01}  + \sum\limits_{j \in {\cal{V}}\left( k \right)} {h_{j,k}^2 P_j^{Up}} } +1\right\}\nonumber\\
&\mathop  \le \limits^{\left( e \right)} c_{00}\left\{ {\sum\limits_{l,j \in {\cal{V}}\left( k \right),l \ne j} {\beta _j
h_{j,k} \beta _l h_{l,k} } } \right.\left[ {\sqrt {E\left( {w_j^2 } \right)E\left( {w_l^2 } \right)} }
\right.\nonumber\\
&\left. { + f_{S,j} f_{S,l} P_S } \right]\left. { + \sum\limits_{j \in {\cal{V}}\left( k \right)} {h_{j,k}^2 P_j^{Up} } } +1 \right\}\nonumber
\end{align}
\begin{align}
&\mathop  \le \limits^{\left( f \right)} c_{00}\left\{ {\sum\limits_{j \in {\cal{V}}\left( k \right)} {h_{j,k}^2 P_j^{Up} }  +
\sum\limits_{l,j \in {\cal{V}}\left( k \right),\atop
  \scriptstyle l \ne j} {h_{j,k} h_{l,k} } } \right.\left. { \sqrt {P_j^{Up} P_l^{Up} } }+1 \right\} \nonumber\\
&=P_k^{Up}
\end{align}
where $c_{00}  = \displaystyle \frac{{P_k^{Up} }}{{\left( {1 + \delta _k } \right)P_{R,k} }}$,
\[
c_{01}=\sum\limits_{l,j \in {\cal{V}}\left( k \right) \atop
  \scriptstyle l \ne j} {\beta _j h_{j,k} \beta _l h_{l,k} \left[ {E\left( {w_j w_l }
\right) + f_{S,j} f_{S,l} P_S } \right]},
\]and

(a)	follows from (6), (7)and (11),

(b)	follows from (10),

(c)	follows from (6),

(d)	follows from the assumption (13),

(e)	follows from the Schwarz inequality,

(f)  also follows from (13), and the fact that\\*$\left[ {\sqrt {E\left( {w_j^2 } \right)E\left( {w_l^2 } \right)} +
f_{S,j} f_{S,l} P_S} \right]^2$\[\le \left[ {E\left( {w_j^2 } \right) + f_{S,j}^2 P_S } \right]\left[ {E\left( {w_l^2 }
\right) + f_{S,l}^2 P_S } \right].\]
Then we complete the proof.
\end{IEEEproof}

\section{Proof of Theorem 1}
\begin{IEEEproof}
We denote the SNR function at the destination of an ideal network where only the nodes at layer $l_0$, $l_0=1,2 \cdots L$, introduce the Gaussian noises as $SNR_{l_0}$. With the identical ANC scheme,
it is obvious that
\begin{equation}
SNR_{l_0}\ge SNR.
\end{equation}
It concludes that the maximum of $SNR_{l_0}$ is no less than the maximum of $SNR$.
We first point out that
\begin{equation}
{\bm{\gamma }} = {\bf{B}}_{l_0} {\bf{g}} = {\bm{G\beta }}_{l_0},
\end{equation}
where
\[{\bm{\beta }}_{l_0}  = \left[ {\beta _1  \cdots \beta _j  \cdots \beta _{n_{l_0} } } \right]^T ,j \in {\cal{L}}_{l_0}
,\]
\[{\bf{B}}_{l_0}  ={\rm{diag}}\left\{ {\beta_1 \cdots \beta {}_j \cdots \beta {}_{n_{l_0} }} \right\},j \in {\cal{L}}_{l_0},\]
\[{\bf{g}} = \left(
{{\bf{h}}_{L - 1}^T {\bf{B}}_{L - 1}  \cdots {\bf{H}}_{l_0} } \right)^T  = \left[ {g_1  \cdots g_j  \cdots g_{n_{l_0} } }
\right]^T,\]
\[{\bf{G}} = {\rm{diag}}\left\{ {g_1  \cdots g_j  \cdots g_{n_{l_0} } } \right\}.
\]
The received signal vector at layer $l_0$ is
\begin{equation}
{\bf{y}}_{l_0}=\left[f_{S,1} \cdots f_{S,j} \cdots f_{S,n_{l_0}}\right]^T x_S+ {\bf{z}}_{l_0}
\end{equation}
and denote by
\begin{equation}
{\bf{P}}_{l_0}=\left[f_{S,1} \cdots f_{S,j} \cdots f_{S,n_{l_0}}\right]^T \sqrt{P_S}
\end{equation}
A solution to
\begin{equation}
\max {\rm{  }}SNR_{l_0}
\end{equation}
can be found by representing the object function in the form of Rayleigh quotient that would then be maximized.
With the notations defined above, $SNR_{l_0}$ can be expressed as
\begin{equation}
SNR_{l_0} \left( {\bm{\gamma }} \right) = \frac{{{\bm{\gamma }}^T {\bf{P}}_{l_0} {\bf{P}}_{l_0}^T {\bm{\gamma }}}}{{{\bm{\gamma }}^T
{\bm{\gamma }}}}.
\end{equation}
The optimal solution is
\begin{equation}
{\bm{\gamma }}_{opt}=c{\bf{P}}_{l_0},
\end{equation}
where $c$ is a constant which is chosen such that the power constraints are satisfied, the corresponding optimal value is
\begin{equation}
SNR_{l_0} \left( {{\bm{\gamma }}_{opt} } \right) = {\bf{P}}_{l_0}^T {\bf{P}}_{l_0},
\end{equation}
which can be interpreted as the total received signal power at nodes in layer $l_0$.

It follows that the larger the elements in ${\bf{P}}_{l_0}$, the larger the optimal value.
From (9) in definition 3, we know that the received signal power at node $k$ in layer $l_0$ is upper bounded by $P_{R,k}$. Therefore the total received signal power at layer $l_0$ is upper bounded by ${\bf{P}}_{R,l_0}^T {\bf{P}}_{R,l_0}$. Consequently, the upper bound of $SNR_{l_0} \left( {{\bm{\gamma }}_{opt} } \right)$ can be given,
\begin{equation}
SNR_{l_0} ^{ Up}\left( {{\bm{\gamma }}_{opt} } \right) = {\bf{P}}_{R,l_0}^T {\bf{P}}_{R,l_0},
\end{equation}
and the corresponding Gaussian channel capacity evaluates to
\begin{equation}
R_{l_0}^{Up}  = \frac{1}{2}\log \left( {1 + {\bf{P}}_{R,l_0}^T {\bf{P}}_{R,l_0} } \right),
\end{equation}
which can be served as an upper bound to the optimal achievable rate of the original network. By taking $l_0=1,2 \cdots L$, a set of upper bounds can be derived which leads to the result
\begin{equation}
R^{Up}  = \mathop {\min }\limits_{l_0  = 1,2 \cdots L} R_{l_0 }^{Up}  = \mathop {\min }\limits_{l_0  = 1,2 \cdots L} \frac{1}{2}\log \left( {1 + {\bf{P}}_{R,l_0 }^T {\bf{P}}_{R,l_0 } } \right)
\end{equation}
such that $R_{opt} \le R^{Up}$ holds.

Then we complete the proof.
\end{IEEEproof}

\section{Proof of Theorem 2}
\begin{IEEEproof}
It is obvious that the maximum achievable rate $R_{opt}$ is no less than the rate derived from the scheme proposed in this paper. To present the results, ${\cal{C}}(x)$ is used to denote the capacity of Gaussian channel with SNR $x$.
\begin{align}
R_{opt}& \ge {\cal{C}}\left( {\frac{{E\left[ {\left( {{\bf{h}}_{L - 1}^T {\bf{B}}_{L - 1}  \cdots {\bf{H}}_1 {\bf{B}}_1 {\bf{h}}_0 x_s } \right)^2 } \right]}}{{E\left[ {\left( {\sum\limits_{l = 1}^{L - 1} {{\bf{h}}_{L - 1}^T {\bf{B}}_{L - 1}  \cdots {\bf{H}}_l {\bf{B}}_l {\bf{z}}_l } } \right)^2 } \right] + E\left[ {z_D^2 } \right]}}} \right)\nonumber\\
&\mathop  = \limits^{(a)} {\cal{C}}\left( {\frac{{\displaystyle
 \frac{{c_1 ^2 \left( {{\bf{P}}_{R,l_0 }^T {\bf{P}}_{R,l_0 } } \right)^2 }}{{\displaystyle\left( {1 + \delta_0 } \right)^{l_0  - 1} }}}}{{c_4+ c_1 ^2 {\bf{P}}_{R,l_0 }^T {\bf{P}}_{R,l_0 }  + 1}}} \right)\nonumber\\
&\mathop  \ge \limits^{(b)} {\cal{C}}\left( { \frac{{\displaystyle \frac{{{\bf{P}}_{R,l_0 }^T {\bf{P}}_{R,l_0 } }}{{\left( {1 + \delta_0 } \right)^{l_0  - 1} }}}}{{\left[ {1 - \frac{1}{{\left( {1 + \delta_0 } \right)^{l_0  - 1} }}} \right]{\bf{P}}_{R,l_0 }^T {\bf{P}}_{R,l_0 }  + c_3 }}} \right),
\end{align}
where $c_4  = E\left[ {\left( {\sum\limits_{l = 1}^{l_0  - 1} {{\bf{h}}_{L - 1}^T {\bf{B}}_{L - 1}  \cdots {\bf{H}}_l {\bf{B}}_l {\bf{z}}_l } } \right)^2 } \right] + c_2$.

We first claim that $(a)$ follows from
\begin{align}
&E\left[ \left( {{\bf{h}}_{L - 1}^T {\bf{B}}_{L - 1}  \cdots {\bf{H}}_{l_0} {\bf{B}}_{l_0} {\bf{z}}_{l_0} } \right)^2  \right] = E\left[{\bf{g}}^T{\bf{B}}_{l_0}{\bf{z}}_{l_0}{\bf{z}}_{l_0}^T{\bf{B}}_{l_0}{\bf{g}}\right]\nonumber\\
&= E\left[{\bf{g}}^T{\bf{B}}_{l_0}{\bf{z}}_{l_0}{\bf{z}}_{l_0}^T{\bf{B}}_{l_0}{\bf{g}}\right]= {\bm{\beta}}^T_{l_0} {\bf{G}}{\bf{G}}{\bm{\beta}}_{l_0}= c_1^2 {\bf{P}}_{R,l_0 }^T {\bf{P}}_{R,l_0 },\nonumber
\end{align}
where $\bf{g}$ and $\bf{G}$ are defined in (77) and the last equality follows from (34), and
\begin{align}
E & \left[ ( {{\bf{h}}_{L - 1}^T {\bf{B}}_{L - 1}  \cdots {\bf{H}}_1 {\bf{B}}_1 {\bf{h}}_0 x_s )^2 } \right] \nonumber\\
& = \frac{1}{{\left( {1 + \delta _0 } \right)^{l_0  - 1} }}{\left( {{\bf{h}}_{L - 1}^T {\bf{B}}_{L - 1}  \cdots {\bf{H}}_{l_0 } {\bf{B}}_{l_0 } {\bf{P}}_{R,l_0 } } \right)^2 } \nonumber\\
& = \frac{{\left(c_1 {\bf{P}}_{R,l_0 }^T {\bf{P}}_{R,l_0 } \right)^2}}{{\left( {1 + \delta _0 } \right)^{l_0  - 1} }}. \nonumber
\end{align}

Then, we claim that $(b)$ follows from
\begin{align}
E&\left[ {\left( {\sum\limits_{l = 1}^{l_0  - 1} {{\bf{h}}_{L - 1}^T {\bf{B}}_{L - 1}  \cdots {\bf{H}}_l {\bf{B}}_l {\bf{z}}_l } } \right)^2 } \right] \nonumber\\
& = \sum\limits_{l = 1}^{l_0  - 1} {E\left[ {\left( {{\bf{h}}_{L - 1}^T {\bf{B}}_{L - 1}  \cdots {\bf{H}}_l {\bf{B}}_l {\bf{z}}_l } \right)^2 } \right]} \nonumber\\
& \le \sum\limits_{d = 1}^{l_0  - 1} {\frac{{\delta _0 }}{{\left( {1 + \delta _0 } \right)^d }} {\left( {{\bf{h}}_{L - 1}^T {\bf{B}}_{L - 1}  \cdots {\bf{H}}_{l_0 } {\bf{B}}_{l_0 } {\bf{P}}_{R,l_0 } } \right)^2 } }  \nonumber\\
& = \sum\limits_{d = 1}^{l_0  - 1} {\frac{{\delta _0 }}{{\left( {1 + \delta _0 } \right)^d }}c_1 ^2 \left( {{\bf{P}}_{R,l_0 }^T {\bf{P}}_{R,l_0 } } \right)^2 },  \nonumber
\end{align}
where the inequality follows from Lemma 1 in [15].

By taking $l_0=1,2 \cdots L$, a set of lower bounds can be derived which leads to the result.
\end{IEEEproof}

\section{Proof of Lemma 2}
\begin{IEEEproof}
To prove Lemma $2$, we justify the following proposition. To have each rate set ${\cal{R}}^{out}_1({\bf{B}})$, it is sufficient to take ${\bf{x}}={\bf{B}\bf{h}}_1$ in the linear subspace $span\left\{ {{\bf{h}}_{0i} ,i = 1,2} \right\}$. Then Lemma 2 can be viewed as an directed corollary of such proposition.

On one hand, any n-dimensional vector $\bf{x}$ can be decomposed as
\setcounter{equation}{87}
\begin{equation}
{\bf{x}}={\bf{x}}_1+{\bf{x}}_2,
\end{equation}
where ${\bf{x}_1} \in span\left\{ {{\bf{h}}_{0i} ,i = 1,2} \right\}$, ${\bf{x}_2} \in span^{\perp}\left\{ {{\bf{h}}_{0i} ,i = 1,2} \right\}$, and thus ${\bf{x}}_1^{T}{\bf{x}}_2=0$.

On the other hand, $span^{\perp}\left\{ {{\bf{h}}_{0i} ,i = 1,2} \right\}$ is the solution space of the homogenous linear equations ${\bf{H}}_0^T {\bf{x}}={\bf{0}}$. Therefore, it follows that ${\bf{h}}_{0i}^{T}{\bf{x}}_2=0$, $i=1,2$.

Therefore, from (52), it is easy to verify that given any amplification vector ${\bf{x}}={\bf{B}\bf{h}}_1$, there exists another one ${\bf{y}}={\bf{B}}'{\bf{h}}_1=c{\bf{x}_1}$ such that ${\cal{R}}^{out}_1({\bf{B}})\subset {\cal{R}}^{out}_1({\bf{B}}')$. Note that the constant $c$ ensures the scheme ${\bf{B}}'$ in the feasible set $\left\{{\bf{B}}\right\}$ and has no effect on the rate set ${\cal{R}}^{out}_1({\bf{B}}')$.

In other words, the corresponding vector component of the amplification vector belonging to $span^{\perp} \left\{ {{\bf{h}}_{0i} ,i = 1,2} \right\}$  will not contribute to the received SNR for the three Gaussian capacities in the rate set (52).

Then we complete the proof.
\end{IEEEproof}

\section{Proof of Lemma 3}
\begin{IEEEproof}
The lemma can be proved by a consequence of careful calculation. Several basic conditions are described first. As assumed in the previous section that all the channel gains are positive values, the angles $\alpha$ and $\beta$ are both in $\left[ {0,\frac{\pi }{2}} \right]$. Without loss of generality, assume $0 \le \alpha  \le \beta  \le \frac{\pi }{2}$ hence $0 \le \beta  - \alpha  \le \frac{\pi }{2}$. From (59), we can see that the outer bound ${\cal{R}}^{out}_1{(\theta)}$ only dependents on the inner product of the channel vectors and the amplification vector. Although $\theta$ can take all values in $\left[ { - \pi ,\pi } \right]$ , it is easy to find that ${\bf{x}}^T {\bf{h}}_{01}  = \cos ^2 \left( {\theta - \alpha } \right) = \cos ^2 \left( {\theta - \alpha  \pm \pi } \right)$ and the same argument applies to the inner product between $\bf{x}$ and $\bf{h}_{02}$. This implies that the value of $\theta$ can be limited to $\left[ { - \frac{\pi }{2},\frac{\pi }{2}} \right]$. Then the proof can be completed by the following discussions.
\begin{itemize}
\item Case $1$. When $\theta \in \left[ {\beta ,\frac{\pi }{2}} \right]$,
\begin{equation}
{\bf{x}}^T {\bf{h}}_{01}  = \cos ^2 \left( {\theta - \alpha } \right)\mathop  \le \limits^{(a)} \cos ^2 \left( {\beta  - \alpha } \right),
\end{equation}
where (a) follows by $0 \le \beta  - \alpha  \le \theta - \alpha  \le \frac{\pi }{2}$, and
\begin{equation}
{\bf{x}}^T {\bf{h}}_{02}  = \cos ^2 \left( {\theta - \beta } \right) \le \cos ^2 \left( {\beta  - \beta } \right) = 1.
\end{equation}
Therefore, by taking $\theta^{*}=\beta$, ${\cal{R}}^{(1)}{(\theta)} \subset {\cal{R}}^{out}_1{(\theta^{*})}$ holds for all $\theta \in \left[ {\beta ,\frac{\pi }{2}} \right]$.

\item Case $2$. When $\theta \in \left[ {\alpha,\beta } \right]$, there is nothing to prove.
\item Case $3$. When $\theta \in \left[ {\beta  - \frac{\pi }{2},\alpha } \right]$,
\begin{equation}
{\bf{x}}^T {\bf{h}}_{01}  = \cos ^2 \left( {\theta - \alpha } \right) \le \cos ^2 \left( {\alpha  - \alpha } \right) = 1,
\end{equation} and
\begin{equation}
{\bf{x}}^T {\bf{h}}_{02}  = \cos ^2 \left( {\beta  - \theta} \right)\mathop  \le \limits^{(a)} \cos ^2 \left( {\beta  - \alpha } \right),
\end{equation}
where (a) follows by  $0 \le \beta  - \alpha  \le \beta  - \theta \le \frac{\pi }{2}$.

Therefore, by taking $\theta^{*}=\alpha$, ${\cal{R}}^{out}_1{(\theta)} \subset {\cal{R}}^{out}_1{(\theta^{*})}$ holds for all $\theta \in \left[ {\beta  - \frac{\pi }{2},\alpha } \right]$.

\item Case $4$. When  $\theta \in \left[ {\alpha  - \frac{\pi }{2},\beta  - \frac{\pi }{2}} \right]$,

let $\Delta  = \beta  - \frac{\pi }{2} - \theta \in \left[ {0,\beta  - \alpha } \right]$ and $\theta' = \beta  - \frac{\pi }{2} + \Delta  \in \left[ {\beta  - \frac{\pi }{2},\beta  + \left( {\beta  - \alpha  - \frac{\pi }{2}} \right)} \right]$,
\begin{align}
{\bf{x}}^T {\bf{h}}_{02} & = \cos ^2 \left( {\beta  - \theta} \right) = \cos ^2 \left( {\frac{\pi }{2} + \Delta } \right) \nonumber \\
&= \cos ^2 \left( {\frac{\pi }{2} - \Delta } \right) = \cos ^2 \left( {\beta  - \theta'} \right)
\end{align}
$1)$. If $\theta' \le \alpha$, then
\begin{align}
{\bf{x}}^T {\bf{h}}_{01}  &= \cos ^2 \left( {\alpha  - \theta} \right) = \cos ^2 \left( {\alpha  - \beta  + \frac{\pi }{2} + \Delta } \right) \nonumber \\
& \mathop  \le \limits^{(a)} \cos ^2 \left( {\alpha  - \beta  + \frac{\pi }{2} - \Delta } \right) = \cos ^2 \left( {\alpha  - \theta'} \right),
\end{align}
where (a) follows from $0 \le \alpha  - \theta'={\alpha  - \beta  + \frac{\pi }{2} - \Delta } \le {\alpha  - \beta  + \frac{\pi }{2} + \Delta } = \alpha  - \theta \le \frac{\pi }{2}$.

Therefore, in such case, for any $\theta$ there always exists a $\theta' \in \left[\beta- \frac{\pi}{2},\alpha \right]$ that ${\cal{R}}^{out}_1{(\theta)} \subset {\cal{R}}^{out}_1{(\theta')}$ holds. Hence from the result of case $3$, by taking $\theta^{*}=\alpha$, ${\cal{R}}^{out}_1{(\theta)} \subset {\cal{R}}^{out}_1{(\theta^{*})}$ holds.
\\$2)$. If $\alpha  < \theta' \le \beta$, then
\begin{align}
{\bf{x}}^T {\bf{h}}_{01}  &= \cos ^2 \left( {\alpha  - \theta} \right) = \cos ^2 \left( {\alpha  - \beta  + \frac{\pi }{2} + \Delta } \right) \nonumber \\
&\mathop  \le \limits^{(a)} \cos ^2 \left( {\beta  - \alpha  - \frac{\pi }{2} + \Delta } \right) = \cos ^2 \left( {\theta' - \alpha } \right),
\end{align}
where (a) follows from $0 \le \alpha  - \theta'={\alpha  - \beta  + \frac{\pi }{2} - \Delta } \le {\alpha  - \beta  + \frac{\pi }{2} + \Delta } = \alpha  - \theta \le \frac{\pi }{2}$.

Therefore, in such case, for any $\theta$ there always exists a $\theta' \in \left[\alpha ,\beta \right]$ that ${\cal{R}}^{out}_1{(\theta)} \subset {\cal{R}}^{out}_1{(\theta')}$ holds.
\item Case 5. When $\theta \in \left[ { - \frac{\pi }{2},\alpha  - \frac{\pi }{2}} \right]$,
\begin{align}
{\bf{x}}^T {\bf{h}}_{01} &= \cos ^2 \left( {\theta - \alpha } \right) = \cos ^2 \left( {\theta - \alpha  + \pi } \right) \nonumber \\
& \mathop  \le \limits^{(a)} \cos ^2 \left( {\beta  - \alpha } \right),
\end{align}
where (a) follows from $ \frac{\pi }{2} - \beta \le \frac{\pi }{2} - \alpha \le \theta - \alpha  + \pi \le \frac{\pi }{2} $, and
\begin{equation}
{\bf{x}}^T {\bf{h}}_{02}  = \cos ^2 \left( {\theta - \beta } \right) \le \cos ^2 \left( {\beta  - \beta } \right) = 1.
\end{equation}
Therefore, by taking $\theta^{*}=\beta$, ${\cal{R}}^{out}_1{(\theta)} \subset {\cal{R}}^{out}_1{(\theta^{*})}$ holds for all $\theta \in \left[ { - \frac{\pi }{2},\alpha  - \frac{\pi }{2}} \right]$.
\end{itemize}
Then we complete the proof.
\end{IEEEproof}

\section{Proof of Theorem 5}
\begin{IEEEproof}
As proved in Lemma 2, it is assumed without loss of generality that the amplification vector ${\bf{x}}\in span\{{\bf{h}}_{01},{\bf{h}}_{02}\}$.
Note that in the proof of Lemma 2, there is no restriction on the dimension of ${\bf{x}}$.
To apply the same argument to obtain (56)-(58), it is assumed that ${\bf{x}}$, ${\bf{h}}_{01}$ and
${\bf{h}}_{02}$ are unit vectors. Then we find a set of unit orthogonal bases of linear space
$span\{{\bf{h}}_{01},{\bf{h}}_{02}\}$, denoted by $({\bf{u}}_1, {\bf{u}}_2)$ such that
\setcounter{equation}{97}
\begin{equation}
{\bf{x}}=\cos \theta {\bf{u}}_1+ \sin \theta {\bf{u}}_2 ,
\end{equation}
\begin{equation}
{\bf{h}}_{01}=\cos\alpha{\bf{u}}_1+\sin\alpha{\bf{u}}_2 ,
\end{equation}
\begin{equation}
{\bf{h}}_{02}=\cos\beta{\bf{u}}_1+\sin\beta{\bf{u}}_2 ,
\end{equation}
where $\cos \theta= {\bf{x}}^{T} {\bf{u}}_1$, $\cos \alpha= {\bf{h}}_{01}^{T} {\bf{u}}_1$ and $\cos \beta= {\bf{h}}_{02}^{T} {\bf{u}}_1$,
and $\alpha , \beta \in \left[0,\frac{\pi}{2}\right]$. Since all the channel gains are positive, it is clear that
$|\alpha - \beta|< \frac{\pi}{2}$.

Again, the rate set ${\cal{R}}({\bf{B}})$ for a two-hop MAC with $n$ relay nodes can be recast as (59) with (98)-(100). Therefore, by Lemma 3 it follows
that to have ${\cal{R}}^{out}_1(\left\{\theta\right\})$, it is sufficient to take $\theta \in [\alpha , \beta]$, and by Lemma 4 it follows that when
$\left| {\alpha  - \beta } \right| \le \frac{\pi }{4}$, ${\cal{R}}^{out}_1 \left( \{ \theta \} \right)$ is convex.

Then we complete the proof.
\end{IEEEproof}

\section{Proof of Lemma 8}
\begin{IEEEproof}
\begin{figure}
  \centering
  \includegraphics[width=2.5in]{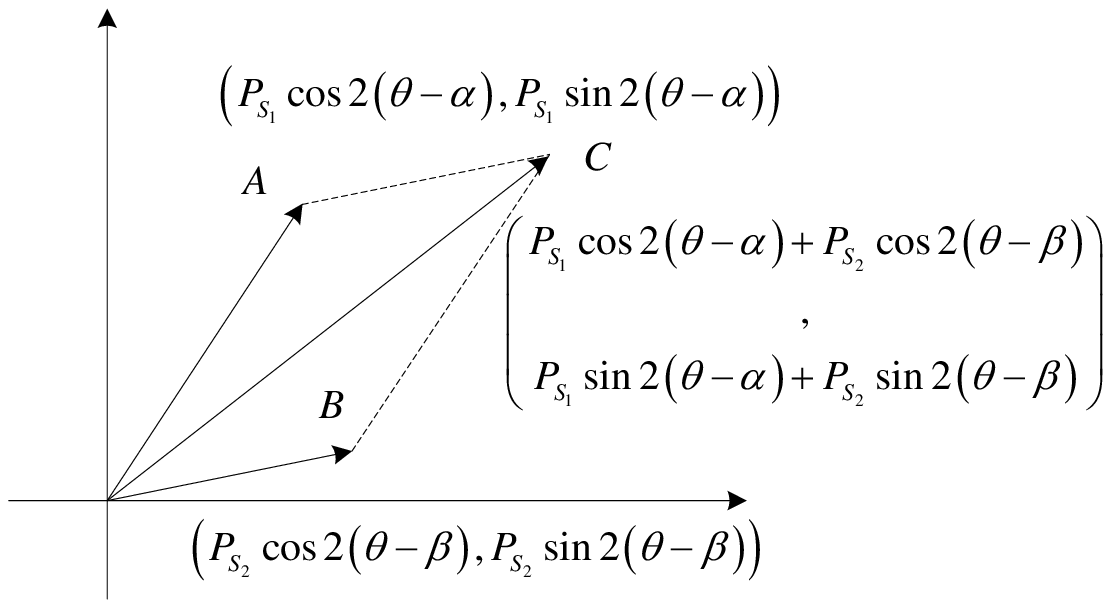}
  \caption{Complex plane representation 1}
\end{figure}
\begin{figure}
  \centering
  \includegraphics[width=2.5in]{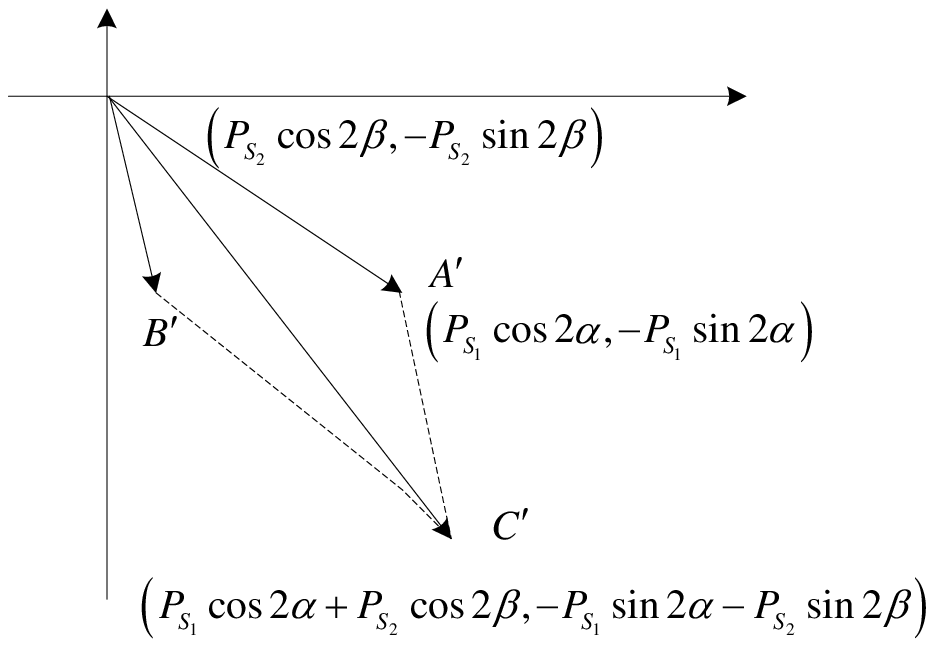}
  \caption{Complex plane representation 2}
\end{figure}
The function of the "upper-diagonal" corner point of each rate set ${\cal{R}}^{out}_1(\theta)$ is given as follows. We consider the boundary points determined by $\theta \in [\theta^{*}(\frac{1}{2}), \beta]$.
\begin{equation}
\left\{
\begin{aligned}
 x &= \frac{1}{2}\ln \left( {1 + \phi \left( \theta  \right)} \right) - \frac{1}{2}\ln \left( {1 + \phi _2 \left( \theta  \right)} \right) \\
 y &= \frac{1}{2}\ln \left( {1 + \phi _2 \left( \theta  \right)} \right) \\
 \end{aligned}
\right.
\end{equation}

\newcounter{TempEqCnt102}
\setcounter{TempEqCnt102}{\value{equation}}
\setcounter{equation}{102}
\begin{figure*}[ht]
\begin{align}
g_2 \left( \theta  \right) =& \phi '\left( \theta  \right)\left( {1 + \phi _2 \left( \theta  \right)} \right) - \phi '_2 \left( \theta  \right)\left( {1 + \phi \left( \theta  \right)} \right) \nonumber\\
= & \left[ { - P_{S_1 } \sin 2\left( {\theta  - \alpha } \right) - P_{S_2 } \sin 2\left( {\theta  - \beta } \right)} \right]\left[ {1 + P_{S_2 } \cos ^2 \left( {\theta  - \beta } \right)} \right]\nonumber\\
&+ P_{S_2 } \sin 2\left( {\theta  - \beta } \right)\left[ {1 + P_{S_1 } \cos ^2 \left( {\theta  - \alpha } \right) + P_{S_2 } \cos ^2 \left( {\theta  - \beta } \right)} \right]\nonumber\\
= & - P_{S_1 } P_{S_2 } \cos ^2 \left( {\theta  - \beta } \right)\sin 2\left( {\theta  - \alpha } \right) - P_{S_2 }^2 \cos ^2 \left( {\theta  - \beta } \right)\sin 2\left( {\theta  - \beta } \right) - P_{S_1 } \sin 2\left( {\theta  - \alpha } \right) - P_{S_2 } \sin 2\left( {\theta  - \beta } \right) \nonumber\\
&+ P_{S_2 } \sin 2\left( {\theta  - \beta } \right) + P_{S_1 } P_{S_2 } \cos ^2 \left( {\theta  - \alpha } \right)\sin 2\left( {\theta  - \beta } \right) + P_{S_2 }^2 \cos ^2 \left( {\theta  - \beta } \right)\sin 2\left( {\theta  - \beta } \right) \nonumber\\
= & - P_{S_1 } P_{S_2 } \cos ^2 \left( {\theta  - \beta } \right)\sin 2\left( {\theta  - \alpha } \right) - P_{S_1 } \sin 2\left( {\theta  - \alpha } \right) + P_{S_1 } P_{S_2 } \cos ^2 \left( {\theta  - \alpha } \right)\sin 2\left( {\theta  - \beta } \right) \nonumber\\
= & - P_{S_1 } P_{S_2 } \cos ^2 \left( {\theta  - \beta } \right)\sin 2\left( {\theta  - \alpha } \right) - P_{S_1 } \sin 2\left( {\theta  - \alpha } \right) - P_{S_1 } P_{S_2 } \cos ^2 \left( {\theta  - \alpha } \right)\sin 2\left( {\beta  - \theta } \right)
\end{align}
\hrulefill
\vspace*{4pt}
\end{figure*}
\setcounter{TempEqCnt102}{\value{equation}}

\newcounter{TempEqCnt103}
\setcounter{TempEqCnt103}{\value{equation}}
\setcounter{equation}{103}
\begin{figure*}[ht]
\begin{align}
 g_3 \left( \theta  \right) =& \phi ''\left( \theta  \right)\phi '_2 \left( \theta  \right) - \phi '\left( \theta  \right)\phi ''_2 \left( \theta  \right) \nonumber\\
  = &\left[ { - 2P_{S_1 } \cos 2\left( {\theta  - \alpha } \right) - 2P_{S_2 } \cos 2\left( {\theta  - \beta } \right)} \right]\left[ { - P_{S_2 } \sin 2\left( {\theta  - \beta } \right)} \right] \nonumber\\
  &- \left[ { - 2P_{S_2 } \cos 2\left( {\theta  - \beta } \right)} \right]\left[ { - P_{S_1 } \sin 2\left( {\theta  - \alpha } \right) - P_{S_2 } \sin 2\left( {\theta  - \beta } \right)} \right] \nonumber\\
  = &2P_{S_1 } P_{S_2 } \sin 2\left( {\theta  - \beta } \right)\cos 2\left( {\theta  - \alpha } \right) + 2P_{S_2 }^2 \cos 2\left( {\theta  - \beta } \right)\sin 2\left( {\theta  - \beta } \right) \nonumber\\
  &- 2P_{S_1 } P_{S_2 } \cos 2\left( {\theta  - \beta } \right)\sin 2\left( {\theta  - \alpha } \right) - 2P_{S_2 }^2 \cos 2\left( {\theta  - \beta } \right)\sin 2\left( {\theta  - \beta } \right) \nonumber\\
  = &2P_{S_1 } P_{S_2 } \left[ {\sin 2\left( {\theta  - \beta } \right)\cos 2\left( {\theta  - \alpha } \right) - \cos 2\left( {\theta  - \beta } \right)\sin 2\left( {\theta  - \alpha } \right)} \right]\nonumber \\
  = &  - 2P_{S_1 } P_{S_2 } \sin 2\left( {\beta  - \alpha } \right).
\end{align}
\hrulefill
\vspace*{4pt}
\end{figure*}

\begin{figure*}[!t]
\normalsize
\begin{align}
 g_4 \left( \theta  \right) =& \phi '\left( \theta  \right)\phi '_2 \left( \theta  \right) \nonumber\\
  = &\left[ { - P_{S_1 } \sin 2\left( {\theta  - \alpha } \right) - P_{S_2 } \sin 2\left( {\theta  - \beta } \right)} \right]\left[ { - P_{S_2 } \sin 2\left( {\theta  - \beta } \right)} \right] \nonumber\\
  =  &- P_{S_2 } \sin 2\left( {\beta  - \theta } \right)\left[ {P_{S_1 } \sin 2\left( {\theta  - \alpha } \right) + P_{S_2 } \sin 2\left( {\theta  - \beta } \right)} \right] \nonumber\\
 \mathop  = \limits^{\left( a \right)} & - P_{S_2 } \sin 2\left( {\beta  - \theta } \right)\left[ {\sqrt {P_{S_1 }^2  + P_{S_2 }^2  + 2P_{S_1 } P_{S_2 } \cos 2\left( {\beta  - \alpha } \right)} \sin 2\left( {\theta  - \theta _0 } \right)} \right] \nonumber\\
  =  &- \sqrt {P_{S_1 }^2  + P_{S_2 }^2  + 2P_{S_1 } P_{S_2 } \cos 2\left( {\beta  - \alpha } \right)} P_{S_2 } \sin 2\left( {\beta  - \theta } \right)\sin 2\left( {\theta  - \theta _0 } \right).
\end{align}
\hrulefill
\vspace*{4pt}
\end{figure*}
\setcounter{equation}{\value{TempEqCnt102}}

\setcounter{equation}{101}
\begin{align}
\frac{{d^2 y}}{{dx^2 }} = & - \frac{{g_3 \left( \theta  \right)\left( {1 + \phi \left( \theta  \right)} \right)\left( {1 + \phi _2 \left( \theta  \right)} \right) - g_2 \left( \theta  \right)g_4 \left( \theta  \right)}}{{g_2 \left( \theta  \right)}} \nonumber\\
&\times \frac{{2\left( {1 + \phi \left( \theta  \right)} \right)\left( {1 + \phi _2 \left( \theta  \right)} \right)}}{{
\left[ {\left( {1 + \phi \left( \theta  \right)} \right)\phi '_2 \left( \theta  \right)} \right]^2
g_1^2 \left( \theta  \right)}},
\end{align}
where
\[
g_1 \left( \theta  \right) = \frac{{\phi '\left( \theta  \right)\left( {1 + \phi _2 \left( \theta  \right)} \right)}}{{\left( {1 + \phi \left( \theta  \right)} \right)\phi '_2 \left( \theta  \right)}} - 1,
\]
\[
g_2 \left( \theta  \right) = \phi '\left( \theta  \right)\left( {1 + \phi _2 \left( \theta  \right)} \right) - \phi '_2 \left( \theta  \right)\left( {1 + \phi \left( \theta  \right)} \right),
\]
\[
g_3 \left( \theta  \right) =  {\phi ''\left( \theta  \right)\phi '_2 \left( \theta  \right) - \phi '\left( \theta  \right)\phi ''_2 \left( \theta  \right)} ,
\]
and
\[
g_4 \left( \theta  \right) = \phi '\left( \theta  \right)\phi '_2 \left( \theta  \right).
\]

\newcounter{TempEqCnt106}
\setcounter{TempEqCnt106}{\value{equation}}
\setcounter{equation}{106}
\begin{figure*}[ht]
\begin{align}
A &= \left( {P_{S_1 } \cos 2\left( {\theta  - \alpha } \right),P_{S_1 } \sin 2\left( {\theta  - \alpha } \right)} \right),\nonumber\\
B &= \left( {P_{S_2 } \cos 2\left( {\theta  - \beta } \right),P_{S_2 } \sin 2\left( {\theta  - \beta } \right)} \right),\\
C &= ( {P_{S_1 } \cos 2\left( {\theta  - \alpha } \right) + P_{S_2 } \cos 2\left( {\theta  - \beta } \right),P_{S_1 } \sin 2\left( {\theta  - \alpha } \right) + P_{S_2 } \sin 2\left( {\theta  - \beta } \right)} )\nonumber.
\end{align}
\hrulefill
\vspace*{4pt}
\end{figure*}
\setcounter{equation}{\value{TempEqCnt106}}

\newcounter{TempEqCnt110}
\setcounter{TempEqCnt110}{\value{equation}}
\setcounter{equation}{110}
\begin{figure*}[ht]
\begin{equation}
\left\{
\begin{aligned}
\varphi \left( {C'} \right) &=  - {\rm{arctan}}\left( {\frac{{P_{S_1 } \sin 2\alpha  + P_{S_2 } \sin 2\beta }}{{P_{S_1 } \cos 2\alpha  + P_{S_2 } \cos 2\beta }}} \right),&\frac{{P_{S_1 } \sin 2\alpha  + P_{S_2 } \sin 2\beta }}{{P_{S_1 } \cos 2\alpha  + P_{S_2 } \cos 2\beta }} \ge 0 \\
 \varphi \left( {C'} \right) &=  - \left[ {\pi  + {\rm{arctan}}\left( {\frac{{P_{S_1 } \sin 2\alpha  + P_{S_2 } \sin 2\beta }}{{P_{S_1 } \cos 2\alpha  + P_{S_2 } \cos 2\beta }}} \right)} \right],&\frac{{P_{S_1 } \sin 2\alpha  + P_{S_2 } \sin 2\beta }}{{P_{S_1 } \cos 2\alpha  + P_{S_2 } \cos 2\beta }} \le 0
\end{aligned}
\right.
\end{equation}
\hrulefill
\vspace*{4pt}
\end{figure*}
\setcounter{equation}{\value{TempEqCnt110}}

To verify the concavity of the parametric function (101), we carefully check each term in (102). The computation of $g_2 \left( \theta \right)$ is given in (103) on the top of the next page. We claim that $g_2 \left( \theta \right) \le 0$ for $\theta  \in \left[ {\theta^{*}(\frac{1}{2}) ,\beta } \right]$. Because
\begin{align}
2&\left( {\theta  - \alpha } \right) \in \left[ {0,\pi } \right] \nonumber \\
  \Rightarrow  &- P_{S_1 } P_{S_2 } \cos ^2 \left( {\theta  - \beta } \right)\sin 2\left( {\theta  - \alpha } \right) \le 0, \nonumber \\
  &- P_{S_1 } \sin 2\left( {\theta  - \alpha } \right) \le 0 .\nonumber \\
2&\left( {\beta  - \theta } \right) \in \left[ {0,\pi } \right] \nonumber \\
\Rightarrow & - P_{S_1 } P_{S_2 } \cos ^2 \left( {\theta  - \alpha } \right)\sin 2\left( {\beta  - \theta } \right) \le 0 . \nonumber
\end{align}
which implies that each term in (103) is non-positive.

The computation of $g_3\left( \theta \right)$ is given on the top of the page (104), we claim that $g_3 \left( \theta \right)$ is a non-positive constant because
\[
2\left( {\beta  - \alpha } \right) \in \left[ {0,\pi } \right] \\
  \Rightarrow  g_3 (\theta) = - 2 P_{S_1 } P_{S_2 } \sin 2\left( \beta  - \alpha \right) \le 0.
\]

The computation of $g_4\left( \theta \right)$ is given on the top of the page (105), where the equality (a) follows from the discussion below.
\setcounter{equation}{105}
\begin{align}
P_{S_1 } \sin 2\left( {\theta  - \alpha } \right) &+ P_{S_2 } \sin 2\left( {\theta  - \beta } \right) \nonumber\\
&= {\mathop{\rm Im}\nolimits} \left[ {P_{S_1 } e^{j2\left( {\theta  - \alpha } \right)}  + P_{S_2 } e^{j2\left( {\theta  - \beta } \right)} } \right]
\end{align}
We use $A$, $B$ and $C$ to denote $P_{S_1 } e^{j2\left( {\theta  - \alpha } \right)}$, $P_{S_2} e^{j2\left( {\theta  - \beta } \right)}$ and $P_{S_1 } e^{j2\left( {\theta  - \alpha } \right)}+ P_{S_2} e^{j2\left( {\theta  - \beta } \right)}$ respectively as shown in Fig. 19 and (107). Hence,

\setcounter{equation}{107}
\begin{align}
P_{S_1 } \sin 2\left( {\theta  - \alpha } \right) &+ P_{S_2 } \sin 2\left( {\theta  - \beta } \right) = \left| C \right|\sin \varphi \left( C \right),
\end{align}
where $
\left| C \right|^2  = P_{S_1 }^2  + P_{S_2 }^2  + 2P_{S_1 } P_{S_2 } \cos 2\left( {\alpha  - \beta } \right)
$
and $\varphi \left( C \right)$ denotes the phase of $C$. Furthermore, to obtain the phase $\varphi \left( C \right)$, we first derive the phase of $C'=Ce^{-j2 \theta}$. In consequence, $\varphi \left( C \right)=\varphi \left( C' \right)+2\theta$. As shown in Fig. 20, the coordinate of $C'$ is given as
\begin{equation}
C' = \left( {P_{S_1 } \cos 2\alpha  + P_{S_2 } \cos 2\beta , - P_{S_1 } \sin 2\alpha  - P_{S_2 } \sin 2\beta } \right).
\end{equation}
Therefore,
\begin{equation}
\tan \left( {\varphi \left( {C'} \right)} \right) =  - \frac{{P_{S_1 } \sin 2\alpha  + P_{S_2 } \sin 2\beta }}{{P_{S_1 } \cos 2\alpha  + P_{S_2 } \cos 2\beta }},
\end{equation}
hence $\varphi \left( {C'} \right) =  - 2\theta ^{*}(\frac{1}{2})$ (compare (111) and (63)). Therefore, we conclude that equality (a) holds. Applying the same argument as we justified $g_2 (\theta) \le 0$ , for $\theta  \in \left[ {\theta ^{*}(\frac{1}{2}) ,\beta } \right]$,
\begin{align}
 &2\left( {\theta  - \theta^{*}(\frac{1}{2}) } \right) \in \left[ {0,\pi } \right] \Rightarrow \sin 2\left( {\theta  - \theta ^{*}(\frac{1}{2}) } \right) \ge 0 , \nonumber\\
 &2\left( {\beta  - \theta } \right) \in \left[ {0,\pi } \right] \Rightarrow \sin 2\left( {\beta  - \theta } \right) \ge 0 , \nonumber
\end{align}
therefore, $g_4 (\theta) \le 0$.

Finally, substituting the above results into (102), we obtain
\[
\frac{d^2y}{dx^2} \le 0,
\]
which implies that the parametric function (101) is concave.

The same argument also applies to the boundary function determined by the "lower-diagonal" points for $\theta \in [\alpha , \theta^{*}(\frac{1}{2})]$. Therefore the proof is omitted.

We complete the proof.
\end{IEEEproof}



\ifCLASSOPTIONcaptionsoff
  \newpage
\fi

\end{document}